\documentstyle{article}
\begin{document}

  \title{Cancellation Mechanism of FCNCs in $S_4\times Z_2$ Flavor Symmetric Extra U(1) Model}
  \author{Y. Daikoku\footnote{E-mail: yasu\_daikoku@yahoo.co.jp} \quad and \quad H. Okada\footnote{E-mail: HOkada@bue.edu.eg} \\
  {\em Institute for Theoretical Physics, Kanazawa University, Kanazawa 920-1192, Japan.}$^*$ \\
  {\em Centre for Theoretical Physics, The British University in Egypt}$^\dagger$}
  \maketitle
\begin{abstract}
 We propose a $E_6$ inspired supersymmetric model with a non-Abelian discrete flavor symmetry;
 $SU(3)_c\times SU(2)_W\times U(1)_Y \times U(1)_X\times S_4\times Z_2$.
 In our scenario, the additional abelian gauge symmetry; $U(1)_X$, not only solves the $\mu$-problem in the minimal Supersymmetric Standard Model(MSSM),
 but also requires new exotic fields which play an important role in solving flavor puzzles.
 If our exotic quarks can be embedded into a $S_4$ triplet, which corresponds to the number of the generation,
 one finds that dangerous proton decay can be well-suppressed.
Hence, it might be expected that the generation structure for lepton and quark in the SM(Standard Model) can be understood as a new system in order to
stabilize the proton in a supersymemtric standard model (SUSY).
 Moreover, due to the nature of the discrete non-Abelian symmetry itself,  Yukawa coupling constants of our model are drastically reduced.

In our previous work, we actually have found much success.
However,
 we also have
to solve Higgs mediated FCNC at tree level, as is often the case
with such extended Higgs models. In this paper, we propose a
promising mechanism which could make cancellation
between Higgs and SUSY contributions.


\end{abstract}

\newpage

\def\vol#1#2#3{{\bf {#1}} ({#2}) {#3}}
\def\NP{Nucl.~Phys. }
\def\PL{Phys.~Lett. }
\def\PR{Phys.~Rev. }
\def\PRP{Phys.~Rep. }
\def\PRL{Phys.~Rev.~Lett. }
\def\PTP{Prog.~Theor.~Phys. }
\def\MPL{Mod.~Phys.~Lett. }
\def\IJMP{Int.~J.~Mod.~Phys. }
\def\JETP{Sov.~Phys.~JETP }
\def\JP{J.~Phys. }
\def\ZP{Z.~Phys. }
\def\EPJ{Eur.~Phys.~J.}
\def\th#1{hep-th/{#1}}
\def\ph#1{hep-ph/{#1}}
\def\hex#1{hep-ex/{#1}}
\def\PAN{Phys.~Atomic.~Nucl.}
\def\AST{Astrophys.~J.}

\def\equ#1{\begin{equation}
#1
\end{equation}
}
\def\nn{\nonumber}
\def\no{\nonumber}
\def\rvec#1{\overrightarrow{#1}}
\def\lvec#1{\overleftarrow{#1}}

\def\2tvec#1#2{
\left(
\begin{array}{c}
#1  \\
#2  \\
\end{array}
\right)}

\def\mat2#1#2#3#4{
\left(
\begin{array}{cc}
#1 & #2 \\
#3 & #4 \\
\end{array}
\right)
}

\def\Mat3#1#2#3#4#5#6#7#8#9{
\left(
\begin{array}{ccc}
#1 & #2 & #3 \\
#4 & #5 & #6 \\
#7 & #8 & #9 \\
\end{array}
\right)
}

\def\3tvec#1#2#3{
\left(
\begin{array}{c}
#1  \\
#2  \\
#3  \\
\end{array}
\right)}

\def\4tvec#1#2#3#4{
\left(
\begin{array}{c}
#1  \\
#2  \\
#3  \\
#4  \\
\end{array}
\right)}

\def\L{\left}
\def\R{\right}

\def\pl{\partial}

\def\lra{\leftrightarrow}

\def\hbar{\hspace{1mm}\bar{}\hspace{-1mm}h}

\def\eqn#1{
\begin{eqnarray}
#1
\end{eqnarray}
}

\def\eqno#1{
\begin{eqnarray*}
#1
\end{eqnarray*}
}

\def\etensor{\epsilon_{\mu\nu\rho\sigma}}

\def\class#1#2#3#4{
\left\{
\begin{array}{ll}
\displaystyle{#1} &
  \qquad #2 \\[0.2cm]
\displaystyle{#3} &
  \qquad #4
\end{array}
\right.
}

\def\classt#1#2#3#4#5#6{
\left\{
\begin{array}{ll}
\displaystyle{#1} &
  \qquad #2 \\[0.2cm]
\displaystyle{#3} &
  \qquad #4 \\[0.2cm]
\displaystyle{#5} &
  \qquad #6
\end{array}
\right.
}

\def\therefore{
\raisebox{.2ex}{.}
\raisebox{1.2ex}{.}
\raisebox{.2ex}{.}
}

\def\g{{\rm g}}

\def\dbar#1{\overline{\overline{#1}}}

\def\dsl#1{#1\hspace{-2.1mm}\slash}

\section{Introduction}

 It is well-known that the standard model 
based on $G_{SM}=SU(3)_c\times SU(2)_W\times U(1)_Y$ gauge symmetry
is a quite promising theory to describe interactions of the particles.

However,
there are unsolved or non-verifiable points enough, 
in particular, the followings are underlying to be clarified:
\begin{enumerate}
\item The electroweak symmetry breaking scale $M_W\sim 10^2$ ${\rm GeV}$ is unnaturally small
in comparison with the fundamental energy scale such as Planck scale $M_P\sim 10^{18}$ {\rm GeV}.
\item  The number of Yukawa coupling constants is too many 
to give predictions of the quark and lepton mass matrices.
\item  There is no understanding about the meaning of generations.
\end{enumerate}




It is believed that the first point 
is solved by introducing SUSY 
\cite{SUSY}, but there is still naturalness problem in the MSSM.
The superpotential of MSSM has $\mu$-term:
\eqn{
\mu H^U H^D.
}
The parameter $\mu$ has to be fine-tuned to $O(1\ {\rm TeV})$ in order to give appropriate electroweak breaking scale,
 but it is unnatural.
This problem is elegantly solved by introducing an additional U(1) gauge symmetry.
This extra U(1) model is proposed in the context of superstring-inspired $E_6$ model \cite{extra-u1}.
In this model, the bare $\mu$-term is forbidden by the new $U(1)_X$ symmetry, but
the trilinear term including $G_{SM}$ singlet superfield $S$ is allowed:
\eqn{
\lambda SH^UH^D.
}
When this singlet field $S$ develops a vacuum expectation value (VEV),
the $U(1)_X$ gauge symmetry is spontaneously broken and an effective
$\mu$-term; $\mu_{{\rm eff}}H^UH^D$, is generated from this term, where $\mu_{{\rm eff}}=\lambda \L<S\R>$ \cite{mu-problem}.

A promising solution for the second point 
is a flavor symmetry
\footnote{The $E_6$ inspired supersymmetric extension of SM with discrete flavor symmetry has been considered by authors \cite{f-extra-u1}.}.
In fact, the flavor symmetry strongly reduces the Yukawa coupling constants.
Here, we introduce a non-Abelian discrete flavor symmetry involved in triplet representations, 
expecting that the number of the generations for lepton and quark is {\it three}.
The triplet representations are contained in several non-Abelian
discrete symmetry groups \cite{review}, for examples, $S_4$
\cite{s4}, $A_4$ \cite{a4}, $T'$ \cite{t-prime}, $\Delta(27)$
\cite{d27} and $\Delta(54)$ \cite{d54}. In our work, we consider
$S_4\times Z_2$.

A promising solution for the third point can be arose by 
the cooperation with the flavor symmetry and supersymmetry.
In the MSSM, the R-parity conserving operators such as $QQQL,
E^cU^cU^cD^c$ induce the proton decay at unacceptable level. But,
in the extra U(1) model, these operators are forbidden by the
additional gauge symmetry. However, since the extra U(1) model has
additional exotic fields, the Yukawa interactions for the exotic
quarks and leptons and quarks reduce proton life time to
unacceptable level, again. With the $S_4$ flavor symmetry, such a
dangerous proton decay is sufficiently suppressed. Hence, it might
be expected that the generation structure can be understood as a
new system in order to stabilize the proton \cite{s4e6}.

Considering the Higgs sector of our model, there is a serious
problem of flavor changing neutral current (FCNC). Multiple Higgs
interactions with leptons and quarks induce too large FCNC, if the
mass scale of Higgs bosons is at $O(TeV)$ region \cite{e6fcnc}. In
this paper, we show Higgs contributions to FCNC may be cancelled
by SUSY FCNC contributions. This cancellation solution softens the
FCNC constraint on Higgs mass. Because the mass bound of Higgs
mass is at $O(TeV)$ region, our model is testable at LHC or future
colliders.

The paper is organized as follows.
In section 2, we explain the basic structure of $S_4$ flavor symmetric extra U(1) model.
We give the superpotential of quark and lepton sector in section 3,
and of Higgs sector in section 4.
In section 5, we discuss the Higgs and SUSY contributions to FCNC.
Finally, we make a brief summary in section 6.
Experimental values of mixing matrices and masses of quarks and leptons are given in appendix,
which are used to test our models.

\section{The Extra U(1) Model with $S_4$ Flavor Symmetry}

\subsection{The Extra U(1) Model}

The basic structure of the extra U(1) model is given as follows.
At high energy scale, the gauge symmetry of model has two extra U(1)s,
which consists maximal subgroup of $E_6$ as $G_2=G_{SM}\times U(1)_X\times U(1)_Z\subset E_6$.
MSSM superfields and additional superfields are embedded in three 27 multiplets of $E_6$ to cancel anomalies,
which is illustrated in Table 1.
The 27 multiplets are decomposed as ${\bf 27}\supset \L\{Q,U^c,E^c,D^c,L,N^c,H^D,g^c,H^U,\R.$
$\L.g,S\R\}$,
where $N^c$ are right-handed neutrinos (RHN), $g$ and $g^c$ are  exotic quarks, and $S$ are  $G_{SM}$ singlets.
We introduce $G_{SM}\times U(1)_X$ singlets $\Phi$ and $\Phi^c$ to break $U(1)_Z$
which prevents the RHNs from having Majorana mass terms.
If the $G_{SM}\times U(1)_X$ singlets develop the intermediate scale VEVs along the D-flat direction
of $\L<\Phi\R>=\L<\Phi^c\R>$, then the $U(1)_Z$ is broken and the RHNs obtain the mass terms through
the trilinear terms $Y^M\Phi N^cN^c$ in the superpotential. After the symmetry is broken, as the R-parity symmetry
\eqn{
R=\exp\L[\frac{i\pi}{20}(3x-8y+15z)\R]
}
remains unbroken, $G_1=G_{SM}\times U(1)_X\times R$ survives at low energy.
This is the symmetry of the low energy extra U(1) model.

Within the renormalizable operators, the full $G_2$ symmetric superpotential is given as follows:
\eqn{
W_1&=&W_0+W_S+W_B, \\
W_0&=&Y^UH^UQU^c+Y^DH^DQD^c+Y^EH^DLE^c+Y^N H^ULN^c+Y^M\Phi N^cN^c, \\
W_S&=&kSgg^c+\lambda SH^UH^D, \\
W_B&=&\lambda_1 QQg+\lambda_2 g^cU^cD^c+\lambda_3 gE^cU^c+\lambda_4 g^cLQ+\lambda_5gD^cN^c.
}
For simplicity, we drop gauge and generation indices.
Where $W_0$ is the same as the superpotential of the MSSM with the RHNs besides the absence of $\mu$-term, and
$W_S$ and $W_B$ are the new interactions.
In $W_S$, $kSgg^c$ drives the soft SUSY breaking
scalar squared mass of S to negative through the renormalization group equations (RGEs) and
then breaks $U(1)_X$ and generates mass terms of exotic quarks, and $\lambda SH^UH^D$ is source of the effective $\mu$-term.
Therefore, $W_0$ and $W_S$ are phenomenologically necessary.
In contrast, $W_B$ breaks baryon number and leads to very rapid proton decay, which are phenomenologically
unacceptable, so this must be forbidden.

\begin{table}[htbp]
\begin{center}
\begin{tabular}{|c|c|c|c|c|c|c|c|c|c|c|c||c|c|}
\hline
         &$Q$ &$U^c$    &$E^c$&$D^c$    &$L$ &$N^c$&$H^D$&$g^c$    &$H^U$&$g$ &$S$ &$\Phi$&$\Phi^c$\\ \hline
$SU(3)_c$&$3$ &$3^*$    &$1$  &$3^*$    &$1$ &$1$  &$1$  &$3^*$    &$1$  &$3$ &$1$ &$1$   &$1$     \\ \hline
$SU(2)_W$&$2$ &$1$      &$1$  &$1$      &$2$ &$1$  &$2$  &$1$      &$2$  &$1$ &$1$ &$1$   &$1$     \\ \hline
$y=6Y$   &$1$ &$-4$     &$6$  &$2$      &$-3$&$0$  &$-3$ &$2$      &$3$  &$-2$&$0$ &$0$   &$0$     \\ \hline
$x$      &$1$ &$1$      &$1$  &$2$      &$2$ &$0$  &$-3$ &$-3$     &$-2$ &$-2$&$5$ &$0$   &$0$     \\ \hline
$z$      &$-1$&$-1$     &$-1$ &$2$      &$2$ &$-4$ &$-1$ &$-1$     &$2$  &$2$ &$-1$&$8$   &$-8$    \\ \hline
$R$      &$-$ &$-$      &$-$  &$-$      &$-$ &$-$  &$+$  &$+$      &$+$  &$+$ &$+$ &$+$   &$+$     \\ \hline
\end{tabular}
\end{center}
\caption{$G_2$ assignment of fields.
Where the $x$, $y$ and $z$ are charges of $U(1)_X$, $U(1)_Y$ and $U(1)_Z$, and $Y$ is hypercharge.}
\end{table}

\subsection{$S_4$ Flavor Symmetry}

We show how the $S_4$ flavor symmetry forbids the baryon number
violating superpotential $W_B$. Non-Abelian group $S_4$ has two
singlet representations ${\bf 1}$, ${\bf 1'}$, one doublet
representation ${\bf 2}$ and two triplet representations ${\bf
3}$, ${\bf 3'}$, where ${\bf 1}$ is the trivial representation. As
the generation number of quarks and leptons is three, at least one
superfield of $\L\{Q,U^c,E^c,D^c,L,N^c,H^D,g^c,H^U,g,S\R\}$ must
be assigned to triplet of $S_4$ in order to solve flavor puzzle.
As we assume that full $E_6$ symmetry does not realize at Planck
scale, there is no need to assign all superfields to the same
$S_4$ representations. The multiplication rules of these
representations are as follows: \eqn{
\begin{tabular}{lcl}
${\bf 3}\times {\bf 3}={\bf 1}+{\bf 2}+{\bf 3}+{\bf 3'}$, & &
${\bf 3'}\times {\bf 3'}={\bf 1}+{\bf 2}+{\bf 3}+{\bf 3'}$, \\
${\bf 3}\times {\bf 3'}={\bf 1'}+{\bf 2}+{\bf 3}+{\bf 3'}$, & &
${\bf 2}\times {\bf 3}={\bf 3}+{\bf 3'}$, \\
${\bf 2}\times {\bf 3'}={\bf 3}+{\bf 3'}$, & &
${\bf 2}\times {\bf 2}={\bf 1}+{\bf 1'}+{\bf 2}$, \\
${\bf 1'}\times {\bf 3}={\bf 3'}$, & &
${\bf 1'}\times {\bf 3'}={\bf 3}$, \\
${\bf 1'}\times {\bf 2}={\bf 2}$, & &
${\bf 1'}\times {\bf 1'}={\bf 1}$.
\end{tabular}
}
With these rules, it is easily shown that
all the $S_4$ invariants consist of two or three non-trivial representations are given by
\eqn{
&&{\bf 1'}\cdot{\bf 1'},\quad {\bf 2}\cdot{\bf 2},\quad {\bf 3}\cdot{\bf 3},\quad {\bf 3'}\cdot{\bf 3'},\quad
{\bf 1'}\cdot{\bf 2}\cdot{\bf 2},\quad {\bf 1'}\cdot{\bf 3}\cdot{\bf 3'},\quad
{\bf 2}\cdot{\bf 2}\cdot{\bf 2}, \quad
{\bf 2}\cdot{\bf 3}\cdot{\bf 3}, \no \\
&&{\bf 2}\cdot{\bf 3}\cdot{\bf 3'},\quad
{\bf 2}\cdot{\bf 3'}\cdot{\bf 3'},\quad {\bf 3}\cdot{\bf 3}\cdot{\bf 3},\quad
{\bf 3}\cdot{\bf 3}\cdot{\bf 3'},\quad {\bf 3}\cdot{\bf 3'}\cdot{\bf 3'},\quad
{\bf 3'}\cdot{\bf 3'}\cdot{\bf 3'}.
}
From these, one can see that there is no invariant including only one triplet
\footnote{$T'$ does not have this property but $A_4$, $\Delta(27)$ and $\Delta(54)$ have.}.
Therefore, if $g$ and $g^c$ are assigned to triplets and the others are assigned to singlets
or doublets, then $W_B$ is forbidden. This provides a solution to the proton life time problem.

\subsection{Exotic Quark Decay and Proton Decay Suppression}

The absence of $W_B$ makes exotic quarks and proton stable, but
the existence of exotic quarks which have life time longer than
0.1 second spoils the success of Big Ban nucleosynthesis. In order
to evade this problem, the $S_4$ symmetry must be broken.
Therefore, it is assumed that the $S_4$ breaking terms are induced
from non-renormalizable terms. We introduce $G_2$ singlet $T$ as
triplet of $S_4$ and add the quartic terms: \eqn{
W_{NRB}=\frac{1}{M_P}T\L(QQg+g^cU^cD^c+gE^cU^c+g^cLQ+gD^cN^c\R). }
Where the order one coefficients in front of each terms are
omitted for simplicity. When $T$ develops VEV with \eqn{
\frac{\L<T\R>}{M_P}\sim 10^{-12},    \label{condition} } the
phenomenological constraints on the life times of proton and
exotic quarks are satisfied at the same time \cite{f-extra-u1}.
The violation of $S_4$ symmetry gives $S_4$ breaking corrections
to effective superpotential through the non-renormalizable terms
which are expressed in the same manner as Eq.(10): \eqn{
W_{NRFV}=\frac{1}{M^2_P}T^2\L(H^UQU^c+H^DQD^c+H^DLE^c+
H^ULN^c+M'N^cN^c+SH^UH^D\R)+\frac{1}{M_P}TSgg^c. }
 Since the above corrections are negligibly small,
the $S_4$ flavor symmetry approximately holds in low energy effective theory.
One finds that the most economical flavon sector is the one which is exchanged $T$ into
superfield-product; $\Phi\Phi^c/M_P$, by embedding $\Phi^c$ to a $S_4$ triplet
(Hereafter, we call $\Phi$ and $\Phi^c$ as flavon which is the trigger of flavor violation.).
In this case, the condition 
of Eq. (\ref{condition}) correspond to the following relation:
\eqn{
\frac{\langle\Phi\rangle \langle\Phi^c\rangle }{M^2_P}\sim 10^{-12},
}
and then the right-handed neutrino mass scale can be predicted as follows:
\eqn{M_R\sim \langle\Phi\rangle\sim 10^{-6}M_P\sim 10^{12}\ {\rm GeV}.}
Hence, by applying the above relation to the measurement of proton and
exotic quarks (In our model, we call exotic quarks as $g$-quark.) life time,
it is expected that one can determine the right-handed neutrino mass scale.

\section{Quark and Lepton Sector}

At first, we define $W_0$ that contributes mass matrices of quarks and leptons.
Although the $S_4$ symmetry reduces the Yukawa coupling constants,
there is still an overabundance of parameters.
In order to reduce the Yukawa coupling constants further,
we extend the flavor symmetry to $S_4\times Z_2$ \cite{lepton}.
In our model, all chiral superfields are assigned to
the representations of $S_4\times Z_2$ as Table 2.
\begin{table}[htbp]
\begin{center}
\begin{tabular}{|c|c|c|c|c|c|c|c|c|c|c|c|c|}
\hline
     &$Q_i$    &$Q_3$    &$U^c_i$  &$U^c_3$  &$E^c_1$  &$E^c_2$  &$E^c_3$   &$D^c_i$  &$D^c_3$  &$L_i$    &$L_3$    &$N^c_i$\\
      \hline
$S_4$&${\bf 2}$&${\bf 1}$&${\bf 2}$&${\bf 1}$&${\bf 1}$&${\bf 1}$&${\bf 1'}$&${\bf 2}$&${\bf 1}$&${\bf 2}$&${\bf 1}$&${\bf 2}$\\
      \hline
$Z_2$&$+$      &$-$      &$-$      &$+$      &$+$      &$-$      &$+$       &$-$      &$+$      &$+$      &$+$      &$+$\\
      \hline
     &$N^c_3$  &$H^D_i$  &$H^D_3$  &$H^U_i$  &$H^U_3$  &$S_i$    &$S_3$    &$g_a$     &$g^c_a$  &$\Phi_i$ &$\Phi_3$ &$\Phi^c_a$\\
      \hline
$S_4$&${\bf 1}$&${\bf 2}$&${\bf 1}$&${\bf 2}$&${\bf 1}$&${\bf 2}$&${\bf 1}$&${\bf 3}$ &${\bf 3}$&${\bf 2}$&${\bf 1}$&${\bf 3}$\\
      \hline
$Z_2$&$+$      &$+$      &$-$      &$+$      &$-$      &$-$      &$+$      &$+$       &$+$      &$-$      &$+$      &$+$\\
      \hline
\end{tabular}
\end{center}
\caption{$S_4\times Z_2$ assignment of superfields
(Where the index $i$ of the $S_4$ doublets runs $i=1,2$,
and the index $a$ of the $S_4$ triplets runs $a=1,2,3$.)}
\end{table}

The superpotential $W_0$ which is consistent with $G_2$ and the symmetries of Table 2 is given by
\eqn{
W_0&=&Y^U_1H^U_3(Q_1U^c_1+Q_2U^c_2)+Y^U_3H^U_3Q_3U^c_3\no \\
&+&Y^U_4Q_3(H^U_1U^c_1+H^U_2U^c_2)+Y^U_5(H^U_1Q_1+H^U_2Q_2)U^c_3\no \\
&+&Y^D_1H^D_3(Q_1D^c_1+Q_2D^c_2)+Y^D_3H^D_3Q_3D^c_3\no \\
&+&Y^D_4Q_3(H^D_1D^c_1+H^D_2D^c_2)+Y^D_5(H^D_1Q_1+H^D_2Q_2)D^c_3\no \\
&+&Y^N_2\L[H^U_1(L_1N^c_2+L_2N^c_1)+H^U_2(L_1N^c_1-L_2N^c_2)\R] \no \\
&+&Y^N_3H^U_3L_3N^c_3+Y^N_4L_3(H^U_1N^c_1+H^U_2N^c_2) \no \\
&+&Y^E_1E^c_1(H^D_1L_1+H^D_2L_2)+Y^E_2E^c_2H^D_3L_3+Y^E_3E^c_3(H^D_1L_2-H^D_2L_1) \no \\
&+&\frac12 Y^M_1\Phi(N^c_1N^c_1+N^c_2N^c_2)+\frac12 Y^M_3\Phi N^c_3N^c_3.
}
There are sixteen complex Yukawa coupling constants in this superpotential.
The twelve phases of these can be absorbed by
redefinition of the five of six quark superfields
$\{Q_i,Q_3,U^c_i,U^c_3,D^c_i,D^c_3\}$ and seven lepton superfields $\{L_i,L_3,E^c_1,E^c_2,E^c_3,N^c_i,N^c_3\}$.
Without loss of generality, we can define
$Y^U_{3,4,5},Y^D_{4,5},Y^N_{2,4},Y^E_{1,2,3},Y^M_{1,3}$ to be real.
We define the phases of complex Yukawa couplings as follows:
\eqn{
Y^U_1=e^{i\alpha}|Y^U_1|,\quad Y^D_1=e^{i\beta}|Y^D_1|,\quad Y^D_3=e^{i\gamma}|Y^D_3|,\quad Y^N_3=e^{i\delta}|Y^N_3|.
}
We write the VEV of the flavon
as
\eqn{
\L<\Phi\R>=V,
}
and the VEVs of the $SU(2)_W$ doublet Higgses as
\eqn{
&&\L<H^U_1\R>=v_u\cos\theta_u,\quad \L<H^U_2\R>=v_u\sin\theta_u,\quad \L<H^U_3\R>=v'_u, \no \\
&&\L<H^D_1\R>=v_d\cos\theta_d,\quad \L<H^D_2\R>=v_d\sin\theta_d,\quad \L<H^D_3\R>=v'_d,
}
where we assume these VEVs are real and the parameters $V, v_{u,d}, v'_{u,d}$ are non-negative and the relation
\eqn{
\sqrt{v^2_u+v'^2_u+v^2_d+v'^2_d}=174\ {\rm GeV}
}
is satisfied. If we define the non-negative mass parameters as follows:
\eqn{
\begin{tabular}{llll}
$M_1=Y^M_1V$,        & $M_3=Y^M_3V$,          &                     &   \\
$m^u_1=|Y^U_1|v'_u$, & $m^u_3=Y^U_3v'_u$,     & $m^u_4=Y^U_4v_u$,   & $m^u_5=Y^U_5v_u$,  \\
$m^d_1=|Y^D_1|v'_d$, & $m^d_3=|Y^D_3|v'_d$,   & $m^d_4=Y^D_4v_d$,   & $m^d_5=Y^D_5v_d$,  \\
$m^\nu_2=Y^N_2v_u$,  & $m^\nu_3=|Y^N_3|v'_u$, & $m^\nu_4=Y^N_4v_u$, &   \\
$m^l_1=Y^E_1v_d$,    & $m^l_2=Y^E_2v'_d$,     & $m^l_3=Y^E_3v_d$,   &
\end{tabular}
}
then the mass matrices of up-type quarks ($M_u$), down-type quarks ($M_d$), charged leptons ($M_l$),
Dirac neutrinos ($M_D$) and Majorana neutrinos ($M_R$) are given by
\eqn{
\begin{tabular}{ll}
$M_u=\Mat3{e^{i\alpha}m^u_1}{0}{m^u_5\cos\theta_u}
{0}{e^{i\alpha}m^u_1}{m^u_5\sin\theta_u}
{m^u_4\cos\theta_u}{m^u_4\sin\theta_u}{m^u_3}$, &
$M_d=\Mat3{e^{i\beta}m^d_1}{0}{m^d_5\cos\theta_d}
{0}{e^{i\beta}m^d_1}{m^d_5\sin\theta_d}
{m^d_4\cos\theta_d}{m^d_4\sin\theta_d}{e^{i\gamma}m^d_3}$,  \\
$M_l=\Mat3{m^l_1\cos\theta_d}{0}{-m^l_3\sin\theta_d}
{m^l_1\sin\theta_d}{0}{m^l_3\cos\theta_d}
{0}{m^l_2}{0}$, &
$M_D=\Mat3{m^\nu_2\sin\theta_u}{m^\nu_2\cos\theta_u}{0}
{m^\nu_2\cos\theta_u}{-m^\nu_2\sin\theta_u}{0}
{m^\nu_4\cos\theta_u}{m^\nu_4\sin\theta_u}{e^{i\delta}m^\nu_3}$,  \\
$M_R=\Mat3{M_1}{0}{0}{0}{M_1}{0}{0}{0}{M_3}$. &
\end{tabular}
}
After the seesaw mechanism, the light neutrino mass matrix is given by
\eqn{
M_\nu&=&M_DM^{-1}_RM^t_D=\Mat3{\rho^2_2}{0}{\rho_2\rho_4\sin2\theta_u}
{0}{\rho^2_2}{\rho_2\rho_4\cos2\theta_u}
{\rho_2\rho_4\sin2\theta_u}{\rho_2\rho_4\cos2\theta_u}{\rho^2_4+e^{2i\delta}\rho^2_3},
}
where
\eqn{
\rho_2=\frac{m^\nu_2}{\sqrt{M_1}},\quad \rho_4=\frac{m^\nu_4}{\sqrt{M_1}},\quad \rho_3=\frac{m^\nu_3}{\sqrt{M_3}}.
}
In the lepton sector, the mass eigenvalues and diagonalization matrix of charged leptons are given by
\eqn{
V^\dagger_{lR}M^t_l V_{lL}&=&diag(m_e,m_\mu, m_\tau)=(m^l_2,m^l_3,m^l_1), \\
V_{lL}&=&\Mat3{0}{-\sin\theta_d}{\cos\theta_d}
{0}{\cos\theta_d}{\sin\theta_d}
{-1}{0}{0}, \\
V_{lR}&=&\Mat3{0}{0}{1}{1}{0}{0}{0}{1}{0},
}
and those of the light neutrinos are given by
\eqn{
V^t_\nu M_\nu V_\nu&=&diag(e^{i(\phi_1-\phi)}m_{\nu_1},e^{i(\phi_2+\phi)}m_{\nu_2},m_{\nu_3}), \\
V_\nu&=&\Mat3{\sin2\theta_u}{-\cos2\theta_u}{0}
{\cos2\theta_u}{\sin2\theta_u}{0}
{0}{0}{1}
\Mat3{-\sin\theta_\nu}{e^{i\phi}\cos\theta_\nu}{0}
{0}{0}{1}
{e^{-i\phi}\cos\theta_\nu}{\sin\theta_\nu}{0},
}
from Eq.(25) and Eq.(28), the Maki-Nakagawa-Sakata (MNS) matrix is given by
\eqn{
V'_{MNS}&=&V^\dagger_{lL}V_\nu P_\nu=\Mat3{-e^{-i\phi}\cos\theta_\nu}{-\sin\theta_\nu}{0}
{-\cos\bar{\theta}\sin\theta_\nu}{e^{i\phi}\cos\bar{\theta}\cos\theta_\nu}{\sin\bar{\theta}}
{-\sin\bar{\theta}\sin\theta_\nu}{e^{i\phi}\sin\bar{\theta}\cos\theta_\nu}{-\cos\bar{\theta}}P_\nu,
}
where
\eqn{
&&\bar{\theta}=\theta_d+2\theta_u, \\
&&P_\nu=diag(e^{-i(\phi_1-\phi)/2},e^{-i(\phi_2+\phi)/2},1).
}
Following ref. \cite{lepton}, we get
\eqn{
\tan^2\theta_\nu&=&\frac{\sqrt{m^2_{\nu_2}-m^2_{\nu_3}\sin^2\phi}-m_{\nu_3}|\cos\phi|}
{\sqrt{m^2_{\nu_1}-m^2_{\nu_3}\sin^2\phi}+m_{\nu_3}|\cos\phi|}, \\
\sin(\phi_1-\phi_2)&=&\frac{m_{\nu_3}\sin\phi}{m_{\nu_1}m_{\nu_2}}
\L[\sqrt{m^2_{\nu_2}-m^2_{\nu_3}\sin^2\phi}+\sqrt{m^2_{\nu_1}-m^2_{\nu_3}\sin^2\phi}\R], \\
\sin(\phi_1-\phi)&=&\frac{\sin\phi}{m_{\nu_1}}
\L[m_{\nu_3}\sqrt{1-\sin^2\phi}+\sqrt{m^2_{\nu_1}-m^2_{\nu_3}\sin^2\phi}\R].
}
After the redefinition of the fields, the MNS matrix is transformed to the standard form in Eq.(106)
where the parameters are given by
\eqn{
\theta_{13}=0,\quad \theta_{12}=\theta_\nu,\quad \theta_{23}=\bar{\theta},\quad
\alpha'=\frac{\phi_1-\phi_2}{2},\quad \beta'=\frac{\phi_1-\phi}{2}.
}
If the neutrino masses have been measured, the two Majorana phases
$\alpha'$ and $\beta'$ would be predicted by Eqs.(32), (33), (34) and (35).
In addition, $\theta_{13}=0$ is predicted, so totally three predictions are given in the lepton sector.

In the quark sector, the mass eigenvalues and diagonalization matrices of quarks are given as follows:
\eqn{
V^\dagger_{uR}M^t_uV_{uL}&=&diag(m_u,m_c,m_t), \\
V_{uL}&=&V_u\Mat3{1}{0}{0}{0}{1}{0}{0}{0}{e^{i\phi_{uL}}}
\Mat3{\cos\theta_{uL}}{0}{\sin\theta_{uL}}{0}{1}{0}{-\sin\theta_{uL}}{0}{\cos\theta_{uL}}S_{12} , \\
V_{uR}&=&V_u\Mat3{1}{0}{0}{0}{1}{0}{0}{0}{e^{i\phi_{uR}}}
\Mat3{\cos\theta_{uR}}{0}{\sin\theta_{uR}}{0}{1}{0}{-\sin\theta_{uR}}{0}{\cos\theta_{uR}}S_{12} , \\
V_u&=&\Mat3{\cos\theta_u}{-\sin\theta_u}{0}{\sin\theta_u}{\cos\theta_u}{0}{0}{0}{1}, \\
m^2_u&=&(m^u_1)^2 , \\
m^2_c&=&\frac12\L[(m^u_1)^2+(m^u_3)^2+(m^u_4)^2+(m^u_5)^2-\mu^2_u\R] , \\
m^2_t&=&\frac12\L[(m^u_1)^2+(m^u_3)^2+(m^u_4)^2+(m^u_5)^2+\mu^2_u\R] , \\
\mu^2_u&=&\sqrt{\L((m^u_3)^2+(m^u_4)^2-(m^u_1)^2-(m^u_5)^2\R)^2+4L^2_u} , \\
L_u&=&\sqrt{(m^u_1m^u_4\cos\alpha+m^u_3m^u_5)^2+(m^u_1m^u_4\sin\alpha)^2} , \\
R_u&=&\sqrt{(m^u_1m^u_5\cos\alpha+m^u_3m^u_4)^2+(m^u_1m^u_5\sin\alpha)^2} , \\
\tan2\theta_{uL}&=&\frac{2L_u}{(m^u_3)^2+(m^u_4)^2-(m^u_1)^2-(m^u_5)^2}, \\
\tan\phi_{uL}&=&\frac{m^u_1m^u_4\sin\alpha}{m^u_1m^u_4\cos\alpha+m^u_3m^u_5}, \\
\tan2\theta_{uR}&=&\frac{2R_u}{(m^u_3)^2+(m^u_5)^2-(m^u_1)^2-(m^u_4)^2}, \\
\tan\phi_{uR}&=&\frac{-m^u_1m^u_5\sin\alpha}{m^u_1m^u_5\cos\alpha+m^u_3m^u_4}, \\
V^\dagger_{dR}M^t_dV_{dL}&=&diag(m_d,m_s,m_b), \\
V_{dL}&=&V_d\Mat3{1}{0}{0}{0}{1}{0}{0}{0}{e^{i\phi_{dL}}}
\Mat3{\cos\theta_{dL}}{0}{\sin\theta_{dL}}{0}{1}{0}{-\sin\theta_{dL}}{0}{\cos\theta_{dL}}S_{12} , \\
V_{dR}&=&V_d\Mat3{1}{0}{0}{0}{1}{0}{0}{0}{e^{i\phi_{dR}}}
\Mat3{\cos\theta_{dR}}{0}{\sin\theta_{dR}}{0}{1}{0}{-\sin\theta_{dR}}{0}{\cos\theta_{dR}}S_{12} , \\
V_d&=&\Mat3{\cos\theta_d}{-\sin\theta_d}{0}{\sin\theta_d}{\cos\theta_d}{0}{0}{0}{1}, \\
m^2_d&=&(m^d_1)^2 , \\
m^2_s&=&\frac12\L[(m^d_1)^2+(m^d_3)^2+(m^d_4)^2+(m^d_5)^2-\mu^2_d\R] , \\
m^2_b&=&\frac12\L[(m^d_1)^2+(m^d_3)^2+(m^d_4)^2+(m^d_5)^2+\mu^2_d\R] , \\
\mu^2_d&=&\sqrt{\L((m^d_3)^2+(m^d_4)^2-(m^d_1)^2-(m^d_5)^2\R)^2+4L^2_d} , \\
L_d&=&\sqrt{(m^d_1m^d_4\cos\beta+m^d_3m^d_5\cos\gamma)^2+(m^d_1m^d_4\sin\beta-m^d_3m^d_5\sin\gamma)^2} , \\
R_d&=&\sqrt{(m^d_1m^d_5\cos\beta+m^d_3m^d_4\cos\gamma)^2+(m^d_1m^d_5\sin\beta-m^d_3m^d_4\sin\gamma)^2} , \\
\tan2\theta_{dL}&=&\frac{2L_d}{(m^d_3)^2+(m^d_4)^2-(m^d_1)^2-(m^d_5)^2} , \\
\tan\phi_{dL}&=&\frac{m^d_1m^d_4\sin\beta-m^d_3m^d_5\sin\gamma}{m^d_1m^d_4\cos\beta+m^d_3m^d_5\cos\gamma} , \\
\tan2\theta_{dR}&=&\frac{2R_d}{(m^d_3)^2+(m^d_5)^2-(m^d_1)^2-(m^d_4)^2} , \\
\tan\phi_{dR}&=&\frac{-m^d_1m^d_5\sin\beta+m^d_3m^d_4\sin\gamma}{m^d_1m^d_5\cos\beta+m^d_3m^d_4\cos\gamma} , \\
S_{12}&=&\Mat3{0}{1}{0}{-1}{0}{0}{0}{0}{1} , } from which the
Cabibbo-Kobayashi-Maskawa (CKM) matrix is given by \eqn{
&&V_{CKM}=V^\dagger_{uL} V_{dL}= \no \\
&&\Mat3{\cos\tilde{\theta}}{-\sin\tilde{\theta}\cos\theta_{dL}}{-\sin\tilde{\theta}\sin\theta_{dL}}
{\sin\tilde{\theta}\cos\theta_{uL}}
{\cos\tilde{\theta}\cos\theta_{uL}\cos\theta_{dL}+e^{i\bar{\phi}}\sin\theta_{uL}\sin\theta_{dL}}
{\cos\tilde{\theta}\cos\theta_{uL}\sin\theta_{dL}-e^{i\bar{\phi}}\sin\theta_{uL}\cos\theta_{dL}}
{\sin\tilde{\theta}\sin\theta_{uL}}
{\cos\tilde{\theta}\sin\theta_{uL}\cos\theta_{dL}-e^{i\bar{\phi}}\cos\theta_{uL}\sin\theta_{dL}}
{\cos\tilde{\theta}\sin\theta_{uL}\sin\theta_{dL}+e^{i\bar{\phi}}\cos\theta_{uL}\cos\theta_{dL}}, \no \\
} where \eqn{ \tilde{\theta}=\theta_d-\theta_u,\quad
\bar{\phi}=\phi_{dL}-\phi_{uL}. } The experimental values of the
matrix elements and Jarlskog invariant in Eq.(172) are reproduced
by putting \eqn{ \tilde{\theta}=13.3^\circ,\quad
\theta_{uL}=2.05^\circ,\quad \theta_{dL}=0.99^\circ,\quad
\bar{\phi}=-83.9^\circ. } In ref. \cite{lepton}, it is assumed
that the VEVs of Higgs $S_3$ doublets are fixed in the direction
of $\theta_u=\theta_d=\frac{\pi}{4}$, which enforces
$\tilde{\theta}=0$ (and predicts the atmospheric neutrino mixing
angle is maximal). This means the Cabibbo angle is zero. In
contrast, there is no such a condition of vacuum directions in
this model.

Due to an overabundance of free parameters, there is no prediction in quark sector.
But we can show that there exist consistent parameter sets.
For example, if we put
\eqn{
\begin{tabular}{llll}
$\alpha=0.00^\circ$, & $\beta=-83.9^\circ$, & $\gamma=83.9^\circ$,  &  \\
$m^u_1=1.28\ {\rm MeV}$,     & $m^u_3=172\ {\rm GeV}$,     & $m^u_4=17.2\ {\rm GeV}$,       & $m^u_5=6.23\ {\rm GeV}$,  \\
$m^d_1=2.91\ {\rm MeV}$,     & $m^d_3=1.94\ {\rm GeV}$,    & $m^d_4=2.14\ {\rm GeV}$,       & $m^d_5=74.2\ {\rm MeV}$,  \\
$m^l_1=1.75\ {\rm GeV}$,     & $m^l_2=487\ {\rm KeV}$,     & $m^l_3=103\ {\rm MeV}$,        &
\end{tabular}
}
then the quark masses in Eq.(171) and the parameters of CKM matrix in Eq.(67) are reproduced.
In this case, unknown mixing angles $\theta_{uR},\theta_{dR}$ and phases $\phi_{uR},\phi_{dR}$ are given by
\eqn{
\theta_{uR}=5.70^\circ,\quad \theta_{dR}=47.8^\circ,\quad \phi_{uR}=\phi_{uL}=0.00^\circ,\quad
\phi_{dR}=-\phi_{dL}=83.9^\circ.
}
These parameters can be expressed by the perturbative Yukawa coupling constants and the VEVs of Higgs fields
through Eq.(20), for example as follows:
\eqn{
\begin{tabular}{llll}
$v_u=41.4\ {\rm GeV}$,           & $v'_u=150\ {\rm GeV}$,
 & $v_d=60.0\ {\rm GeV}$,        & $v'_d=49.5\ {\rm GeV}$,  \\
$\L|Y^U_1\R|=8.53\times 10^{-6}$,& $\L|Y^U_3\R|=1.15$,              & $\L|Y^U_4\R|=0.415$,& $\L|Y^U_5\R|=0.150$, \\
$\L|Y^D_1\R|=5.87\times 10^{-5}$,& $\L|Y^D_3\R|=0.0392$,            & $\L|Y^D_4\R|=0.0357$,& $\L|Y^D_5\R|=1.23\times 10^{-3}$,  \\
$\L|Y^E_1\R|=0.0292$,            & $\L|Y^E_2\R|=9.84\times 10^{-6}$,& $\L|Y^E_3\R|=1.72\times 10^{-3}$. &
\end{tabular}
}
As all the coupling constants of the model are perturbative,
it is consistent that the fundamental energy scale is much larger than the electroweak scale,
which is the base of naturalness problem.


\section{Higgs sector}

Next, we define Higgs potential and solve its minimum condition approximately.
With gauge symmetry in table 1 and flavor symmetry in table 2, superpotential of Higgs sector is given by
\eqn{
W_H &=& \lambda_1S_3 (H^U_1H^D_1 + H^U_2H^D_2) + \lambda_3S_3 H^U_3H^D_3\no\\
&+&  \lambda_4 H^U_3 (S_1H^D_1 + S_2H^D_2) +   \lambda_5 H^D_3
(S_1H^U_1 + S_2H^U_2)\subset W_S, } where one can take, without
any loss of the generalities, $\lambda_{1,3,4,5}$ as real, by
redefinining of four arbitrary fields of $\{S_i, S_3, H^U_i,
H^U_3, H^D_i, H^D_3\}$. However, this superpotential could have
would-be goldstone bosons when all of the Higgs fields acquire
VEVs, because of an accidental $O(2)$ symmetry induced by the
common rotation of the $S_4$ doublet. In order to avoid the
problem, we assume that the flavor symmetry should be explicitly
broken in the soft scalar mass terms, which can play role in
giving the controllable parameters for the direction of the
$SU(2)$ doublet Higgs VEVs.


As the Higgs potential has too many unknown parameters, we make
several assumptions. In the superpotential, we assume the
parameters $\lambda_i$ are hierarchical for examples, as follows:
\eqn{ \lambda_5\ll \lambda_1=0.03\ll \lambda_3=\lambda_4=0.3. }
Then, we can neglect the first and fourth term  in $W_H$. Note
that, too small $\lambda_1$ is not consistent with chargino mass
bound $M_{chargino}>94GeV$, and too large $\lambda_{3,4}$ make
$Y^U_3$ reach Landau pole below $M_P$. With this assumption,
F-term and D-term contribution to Higgs potential is given by
\eqn{ V_{SUSY}&=&\left|\lambda_3H^U_3H^D_3 \right|^2
+\left|\lambda_4H^U_3H^D_1\right|^2+\left|\lambda_4H^U_3H^D_2\right|^2 \no \\
&+&\left|\lambda_3S_3H^D_3+\lambda_4(S_1H^D_1+S_2H^D_2)\right|^2 \no \\
&+&\left|\lambda_3S_3H^U_3\right|^2
+\left|\lambda_4H^U_3S_1\right|^2+\left|\lambda_4H^U_3S_2\right|^2 \no \\
&+&\frac18g^2_2\sum^3_{A=1}\L[(H^U_a)^\dagger\sigma_AH^U_a+(H^D_a)^\dagger\sigma_AH^D_a\R]^2
+\frac18g^2_Y\L[|H^U_a|^2-|H^D_a|^2\R]^2 \no \\
&+&\frac12 g^2_x\L[x_{H^U}|H^U_a|^2+x_{H^D}|H^D_a|^2+x_S|S_a|^2\R]^2,
}
where index $a$ runs $a=1,2,3$, 
and flavor symmetric SUSY breaking terms are given by
\eqn{
V_{SB}&=&m^2_{H^U}(|H^U_1|^2+|H^U_2|^2)-m^2_{H^U_3}|H^U_3|^2
+m^2_{H^D}(|H^D_1|^2+|H^D_2|^2)+m^2_{H^D_3}|H^D_3|^2 \no \\
&+&m^2_S(|S_1|^2+|S_2|^2)-m^2_{S_3}|S_3|^2 \no \\
&-&\left\{\lambda_3A_3S_3H^U_3H^D_3
+\lambda_4A_4H^U_3(S_1H^D_1+S_2H^D_2)+h.c. \right\}, } where all
parameters in $V_{SB}$ can be real in some SUSY breaking scenario,
for example, in the case that A-parameters are induced by gaugino
mass through RGEs, A-parameters become real. In order to avoid
goldstone bosons, we assume flavor violation in soft scalar mass
terms, and add flavor violating terms as follows: \eqn{
V_{SBFB}&=&-m^2_{BH^U}(H^U_3)^\dagger (H^U_1c_{H^U}+H^U_2s_{H^U})
-m^2_{BS}(S_3)^\dagger (S_1c_S+S_2s_S)+h.c. , } where we assume
flavor violation is induced by VEV of $S_4$ doublet $Z_2$ odd
auxiliary field in hidden sector. In this paper, we do not
consider hidden sector which is beyond our paper. With this
assumption, the term
$m^2_{H^D}(H^D_3)^\dagger(H^D_1c_{H^D}+H^D_2s_{H^D})$ should be
included in $V_{SBFB}$. Here we assume this term is approximately
negligible. In this approximation, all parameters of potential
$V=V_{SUSY}+V_{SB}+V_{SBFB}$ are real, because we can remove the
phases of $m^2_{BH^U}$ and $m^2_{BS}$ by field redefinition. After
the redefinition,  three phases of $m^2_{BH^U,BH^D,BS}$ are
transformed into $\lambda_{1,5},m^2_{BH^D}$ which are assumed to
be small and negligible.

From the defined potential above, the potential minimum conditions are given by
\eqn{
0=\frac{\partial V}{\partial H^U_1}/(v_uc_u)
&=&m^2_{H^U}-m^2_{BH^U}c_{H^U}(v'_u/v_uc_u)+g^2_xx_{H^U}x_S(v^2_s+(v'_s)^2) \no \\
&+&\L\{\frac14(g^2_Y+g^2_2)(v^2_u+(v'_u)^2-v^2_d-(v'_d)^2)\R. \no \\
&+&\L. g^2_xx_{H^U}[x_{H^U}(v^2_u+(v'_u)^2)+x_{H^D}(v_d^2+(v'_d)^2)]\R\}, \\
0=\frac{\partial V}{\partial H^U_2}/(v_us_u)
&=&m^2_{H^U}-m^2_{BH^U}s_{H^U}(v'_u/v_us_u)+g^2_xx_{H^U}x_S(v^2_s+(v'_s)^2) \no \\
&+&\L\{\frac14(g^2_Y+g^2_2)(v_u^2+(v'_u)^2-v_d^2-(v'_d)^2) \R. \no \\
&+&\L. g^2_xx_{H^U}[x_{H^U}(v_u^2+(v'_u)^2)+x_{H^D}(v_d^2+(v'_d)^2)]\R\} , \\
0=\frac{\partial V}{\partial H^U_3}/v'_u
&=&-m^2_{H^U_3}-\lambda_3A_3v'_s(v'_d/v'_u)-\lambda_4A_4[v_sc_s(v_dc_d/v'_u)+v_ss_s(v_ds_d/v'_u)] \no \\
&-&m^2_{BH^U}[c_{H^U}(v_uc_u/v'_u)+s_{H^U}(v_us_u/v'_u)] \no \\
&+&\lambda^2_3(v'_s)^2+\lambda^2_4v_s^2+g^2_xx_{H^U}x_S(v^2_s+(v'_s)^2) \no \\
&+&\L\{\lambda^2_3(v'_d)^2+\lambda^2_4v_d^2
+\frac14(g^2_Y+g^2_2)(v_u^2+(v'_u)^2-v_d^2-(v'_d)^2)\R.  \no \\
&+&\L. g^2_xx_{H^U}[x_{H^U}(v_u^2+(v'_u)^2)+x_{H^D}(v_d^2+(v'_d)^2)]\R\}, \\
0=\frac{\partial V}{\partial H^D_1}/(v_dc_d)
&=&m^2_{H^D}-\lambda_4A_4v'_u(v_sc_s/v_dc_d)
+\lambda_4(v_sc_s/v_dc_d)[\lambda_3v'_sv'_d+\lambda_4(v_sc_sv_dc_d+v_ss_sv_ds_d)] \no \\
&+&g^2_xx_{H^D}x_S(v^2_s+(v'_s)^2) \no \\
&+&\L\{\lambda^2_4(v'_u)^2
-\frac14(g^2_Y+g^2_2)(v_u^2+(v'_u)^2-v_d^2-(v'_d)^2)\R. \no \\
&+&\L. g^2_xx_{H^D}[x_{H^U}(v_u^2+(v'_u)^2)+x_{H^D}(v_d^2+(v'_d)^2)] \R\}, \\
0=\frac{\partial V}{\partial H^D_2}/(v_ds_d)
&=&m^2_{H^D}-\lambda_4A_4v'_u(v_ss_s/v_ds_d)
+\lambda_4(v_ss_s/v_ds_d)[\lambda_3v'_sv'_d+\lambda_4(v_sc_sv_dc_d+v_ss_sv_ds_d)] \no \\
&+&g^2_xx_{H^D}x_S(v^2_s+(v'_s)^2) \no \\
&+&\L\{\lambda^2_4(v'_u)^2
-\frac14(g^2_Y+g^2_2)(v_u^2+(v'_u)^2-v_d^2-(v'_d)^2)\R. \no \\
&+&\L. g^2_xx_{H^D}[x_{H^U}(v_u^2+(v'_u)^2)+x_{H^D}(v_d^2+(v'_d)^2)] \R\}, \\
0=\frac{\partial V}{\partial H^D_3}/v'_d
&=&m^2_{H^D_3}-\lambda_3A_3v'_s(v'_u/v'_d)
+\lambda_3(v'_s/v'_d)[\lambda_3v'_sv'_d+\lambda_4(v_sc_sv_dc_d+v_ss_sv_ds_d)] \no \\
&+&g^2_xx_{H^D}x_S(v^2_s+(v'_s)^2) \no \\
&+&\L\{ \lambda^2_3(v'_u)^2
-\frac14(g^2_Y+g^2_2)(v_u^2+(v'_u)^2-v_d^2-(v'_d)^2)\R. \no \\
&+&\L. g^2_xx_{H^D}[x_{H^U}(v_u^2+(v'_u)^2)+x_{H^D}(v_d^2+(v'_d)^2)]\R\} , \\
0=\frac{\partial V}{\partial S_1}/(v_sc_s)
&=&m^2_S-m^2_{BS}(v'_s/v_sc_s)c_S+g^2_xx^2_S(v^2_s+(v'_s)^2) \no \\
&+&\L\{-\lambda_4A_4v'_u(v_dc_d/v_sc_s)+\lambda^2_4(v'_u)^2
+\lambda_4(v_dc_d/v_sc_s)[\lambda_3v'_sv'_d+\lambda_4(v_sc_sv_dc_d+v_ss_sv_ds_d)]\R. \no \\
&+&\L. g^2_xx_S[x_{H^U}(v_u^2+(v'_u)^2)+x_{H^D}(v_d^2+(v'_d)^2)] \R\}, \\
0=\frac{\partial V}{\partial S_2}/(v_ss_s)
&=&m^2_S-m^2_{BS}(v'_s/v_ss_s)s_S+g^2_xx^2_S(v^2_s+(v'_s)^2) \no \\
&+&\L\{-\lambda_4A_4v'_u(v_ds_d/v_ss_s)+\lambda^2_4(v'_u)^2
+\lambda_4(v_ds_d/v_ss_s)[\lambda_3v'_sv'_d+\lambda_4(v_sc_sv_dc_d+v_ss_sv_ds_d)]\R. \no \\
&+&\L. g^2_xx_S[x_{H^U}(v_u^2+(v'_u)^2)+x_{H^D}(v_d^2+(v'_d)^2)] \R\}, \\
0=\frac{\partial V}{\partial S_3}/v'_s
&=&-m^2_{S_3}-m^2_{BS}[(v_sc_s/v'_s)c_S+(v_ss_s/v'_s)s_S]
+g^2_xx^2_S(v^2_s+(v'_s)^2) \no \\
&+&\L\{ -\lambda_3A_3v'_u(v'_d/v'_s)+\lambda^2_3(v'_u)^2
+\lambda_3(v'_d/v'_s)[\lambda_3v'_sv'_d+\lambda_4(v_sc_sv_dc_d+v_ss_sv_ds_d)]\R. \no \\
&+&\L. g^2_xx_S[x_{H^U}(v_u^2+(v'_u)^2)+x_{H^D}(v_d^2+(v'_d)^2)]
\R\}, } where we define the VEVs of the $G_{SM}$ singlet as \eqn{
\L<S_1\R>=v_s\cos\theta_s,\quad \L<S_2\R>=v_s\sin\theta_s,\quad
\L<S_3\R>=v'_s, } and hereafter we neglect the terms in bracket
$\L\{\quad \R\}$, because those are very small (Note that
$v_{u,d},v'_{u,d}\ll v_s,v'_s$.). Solving the Eqs.(76)-(84), we
get \eqn{
&&m^2_{BS}(v'_s/v_s)\L(\frac{c_S}{c_s}-\frac{s_S}{s_s}\R)=0\quad
(\therefore \quad \theta_s=\theta_S), \\
&&m^2_S-m^2_{BS}(v'_s/v_s)+g^2_xx^2_S(v^2_s+(v'_s)^2)=0, \\
&&-m^2_{S_3}-m^2_{BS}(v_s/v'_s)+g^2_xx^2_S(v^2_s+(v'_s)^2)=0, \\
&&\lambda_4\L[\lambda_3v'_sv'_d+\lambda_4v_sv_d(c_sc_d+s_ss_d)
-A_4v'_u\R]\L(\frac{c_s}{c_d}-\frac{s_s}{s_d}\R)\frac{v_s}{v_d}=0 \quad
 (\therefore\quad \theta_d=\theta_s), \\
&&m^2_{H^D}-\lambda_4A_4v_s(v'_u/v_d)+\lambda_3\lambda_4v_sv'_s(v'_d/v_d)
+\lambda^2_4v^2_s+g^2_xx_Sx_{H^D}(v^2_s+(v'_s)^2)=0, \\
&&m^2_{H^D_3}-\lambda_3A_3v'_s(v'_u/v'_d)+\lambda^2_3(v'_s)^2
+\lambda_3\lambda_4v_sv'_s(v_d/v'_d)
+g^2_xx_Sx_{H^D}(v^2_s+(v'_s)^2)=0, \\
&&m^2_{BH^U}(v'_u/v_u)\L(\frac{c_{H^U}}{c_u}-\frac{s_{H^U}}{s_u}\R)=0\quad
(\therefore\quad \theta_u=\theta_{H^U}), \\
&&m^2_{H^U}-m^2_{BH^U}(v'_u/v_u)+g^2_xx_Sx_{H^U}(v^2_s+(v'_s)^2)=0, \\
&&-m^2_{H^U_3}-\lambda_3A_3v'_s(v'_d/v'_u)-\lambda_4A_4v_s(v_d/v'_u)
-m^2_{BH^U}(v_u/v'_u)\no \\
&&\quad +\lambda^2_3(v'_s)^2+\lambda^2_4v^2_s
+g^2_xx_Sx_{H^U}(v^2_s+(v'_s)^2)=0.
}
Using Eqs.(86)-(94), mass matrices of neutral CP-even ($\phi$) and CP-odd ($\rho$) Higgs bosons are
given by
\eqn{
M^2_\phi&=&
\Mat3{M^2_{uu}}{M^2_{ud}}{0}
{M^2_{du}}{M^2_{dd}}{0}
{0}{0}{M^2_{ss}}, \\
M^2_\rho&=&
\Mat3{M^2_{uu}}{-M^2_{ud}}{0}
{-M^2_{du}}{M^2_{dd}}{0}
{0}{0}{\bar{M}^2_{ss}}, \\
M^2_{ss}&=&
\Mat3{m^2_{BS}(v'_s/v_s)+2g^2_xx^2_S(v_sc_s)^2}{2g^2_xx^2_Sv^2_sc_ss_s}{-m^2_{BS}c_s+2g^2_xx^2_Sv_sv'_sc_s}
{2g^2_xx^2_Sv^2_sc_ss_s}{m^2_{BS}(v'_s/v_s)+2g^2_xx^2_S(v_ss_s)^2}{-m^2_{BS}s_s+2g^2_xx^2_Sv_sv'_ss_s}
{-m^2_{BS}c_s+2g^2_xx^2_Sv_sv'_sc_s}{-m^2_{BS}s_s+2g^2_xx^2_Sv_sv'_ss_s}{m^2_{BS}(v_s/v'_s)+2g^2_xx^2_S(v'_s)^2},\no \\
&& \\
\bar{M}^2_{ss}&=&
\Mat3{m^2_{BS}(v'_s/v_s)}{0}{-m^2_{BS}c_s}
{0}{m^2_{BS}(v'_s/v_s)}{-m^2_{BS}s_s}
{-m^2_{BS}c_s}{-m^2_{BS}s_s}{m^2_{BS}(v_s/v'_s)}, \\
M^2_{uu}&=&
\Mat3{m^2_{BH^U}(v'_u/v_u)}{0}{-m^2_{BH^U}c_u}
{0}{m^2_{Bu}(v'_u/v_u)}{-m^2_{BH^U}s_u}
{-m^2_{BH^U}c_u}{-m^2_{BH^U}s_u}{m^2_{BH^U}(v_u/v'_u)+\lambda_3A_3v'_s(v'_d/v'_u)
+\lambda_4A_4v_s(v_d/v'_u)}, \\
M^2_{dd}&=&
\Mat3{-\lambda^2_4(v_ss_s)^2-\lambda_3\lambda_4v_sv'_s(v'_d/v_d)}{\lambda^2_4v^2_sc_ss_s}{\lambda_3\lambda_4v_sv'_sc_s}
{\lambda^2_4v^2_sc_ss_s}{-\lambda^2_4(v_sc_s)^2-\lambda_3\lambda_4v_sv'_s(v'_d/v_d)}{\lambda_3\lambda_4v_sv'_ss_s}
{\lambda_3\lambda_4v_sv'_sc_s}{\lambda_3\lambda_4v_sv'_ss_s}{-\lambda_3\lambda_4v_sv'_s(v_d/v'_d)} \no \\
&+&
\Mat3{\lambda_4A_4v_s(v'_u/v_d)}{0}{0}
{0}{\lambda_4A_4v_s(v'_u/v_d)}{0}
{0}{0}{\lambda_3A_3v'_s(v'_u/v'_d)}, \\
M^2_{ud}&=& \Mat3{0}{0}{0} {0}{0}{0}
{-\lambda_4A_4v_sc_s}{-\lambda_4A_4v_ss_s}{-\lambda_3A_3v'_s}=(M^2_{du})^t,
} where $v_u,v'_u,v_d,v'_d\ll v_s,v'_s$ is assumed and defined as
\eqn{ (H^U_i)^0=\frac{\phi_{u,i}+i\rho_{u,i}}{\sqrt{2}},\quad
(H^D_i)^0=\frac{\phi_{d,i}+i\rho_{d,i}}{\sqrt{2}},\quad
S_i=\frac{\phi_{s,i}+i\rho_{s,i}}{\sqrt{2}}\quad (i=1,2,3). }
Hereafter we do not consider $\phi_{s},\rho_{s}$, because these
fields do not mix with $\phi_{u,d},\rho_{u,d}$ and never
contribute FCNC. Because the mass matrices are partially
diagonalized as follows \eqn{
&&(M^2_{uu})'=V^t_uM^2_{uu}V_u \no \\
&&=
\Mat3{m^2_{BH^U}(v'_u/v_u)}{0}{-m^2_{BH^U}}
{0}{m^2_{BH^U}(v'_u/v_u)}{0}
{-m^2_{BH^U}}{0}{m^2_{BH^U}(v_u/v'_u)+\lambda_3A_3v'_s(v'_d/v'_u)+\lambda_4A_4v_s(v_d/v'_u)}, \\
&&(M^2_{dd})'=V^t_dM^2_{dd}V_d
=\Mat3{\lambda_4v_s[A_4(v'_u/v_d)-\lambda_3v'_s(v'_d/v_d)]}{0}{\lambda_3\lambda_4v_sv'_s}
{0}{M^2_{\phi'_{d,2}}}{0}
{\lambda_3\lambda_4v_sv'_s}{0}{\lambda_3v'_s[A_3(v'_u/v'_d)-\lambda_4v_s(v_d/v'_d)]}, \no \\
&&\quad M^2_{\phi'_{d,2}}=\lambda_4v_s[A_4(v'_u/v_d)-\lambda_4v_s-\lambda_3v'_s(v'_d/v_d)], \\
&&(M^2_{ud})'=V^t_uM^2_{ud}V_d
=\Mat3{0}{0}{0}
{0}{0}{0}
{-\lambda_4A_4v_s}{0}{-\lambda_3A_3v'_s},
}
where $V_u$ and $V_d$ are defined  in Eq. (39) and Eq. (53), respectively,
one can see that the mixed states
\eqn{
&&\phi'_{u,2}=-\phi_{u,1}s_u+\phi_{u,2}c_u,\quad \rho'_{u,2}=-\rho_{u,1}s_u+\rho_{u,2}c_u \\
&&\phi'_{d,2}=-\phi_{d,1}s_d+\phi_{d,2}c_d,\quad
\rho'_{d,2}=-\rho_{d,1}s_d+\rho_{d,2}c_d } are mass eigenstates.
Note that CP-even Higgs bosons $\phi'_{u,2},\phi'_{d,2}$ and
CP-odd Higgs bosons $\rho'_{u,2},\rho'_{d,2}$ have the same mass
eigenvalues in this approximation, respectively.

\section{Cancellation of Higgs and SUSY-FCNC Contributions}

Finally, we evaluate the Higgs and SUSY contributions to FCNC.
Here we calculate $K^0-\bar{K}^0$, $B^0-\bar{B}^0$ and
$D^0-\bar{D}^0$ mass differences.

\subsection{Higgs contributions}

First, we explain how Higgs bosons mediate FCNCs. Yukawa coupling
interactions of quarks and charged leptons between neutral Higgs
bosons are given by \eqn{ -{\cal L}_Y&=&
(\bar{u}_1,\bar{u}_2,\bar{u}_3)_R
\Mat3{Y^U_1(H^U_3)^0}{0}{Y^U_4(H^U_1)^0}
{0}{Y^U_1(H^U_3)^0}{Y^U_4(H^U_2)^0}
{Y^U_5(H^U_1)^0}{Y^U_5(H^U_2)^0}{Y^U_3(H^U_3)^0}
\3tvec{u_1}{u_2}{u_3}_L \no \\
&+&(\bar{d}_1,\bar{d}_2,\bar{d}_3)_R
\Mat3{Y^D_1(H^D_3)^0}{0}{Y^D_4(H^D_1)^0}
{0}{Y^D_1(H^D_3)^0}{Y^D_4(H^D_2)^0}
{Y^D_5(H^D_1)^0}{Y^D_5(H^D_2)^0}{Y^D_3(H^D_3)^0}
\3tvec{d_1}{d_2}{d_3}_L \no \\
&+&(\bar{l}_1,\bar{l}_2,\bar{l}_3)_R
\Mat3{Y^E_1(H^D_1)^0}{Y^E_1(H^D_2)^0}{0} {0}{0}{Y^E_2(H^D_3)^0}
{-Y^E_3(H^D_2)^0}{Y^E_3(H^D_1)^0}{0} \3tvec{l_1}{l_2}{l_3}_L+h.c.
. } With the basis that quark and lepton mass matrices are
diagonal, these terms are rewritten by \eqn{ -{\cal L}_Y&=&
\frac{1}{\sqrt{2}}(\bar{u},\bar{c},\bar{t})_R
\Mat3{Y^U_1(\phi_{u,3}+i\rho_{u,3})}{Y^U_4s_{uL}(\phi'_{u,2}+i\rho'_{u,2})}{-Y^U_4c_{uL}(\phi'_{u,2}+i\rho'_{u,2})}
{Y^U_5s_{uR}(\phi'_{u,2}+i\rho'_{u,2})}{H^U_{22}}{H^U_{23}}
{-Y^U_5c_{uR}(\phi'_{u,2}+i\rho'_{u,2})}{H^U_{32}}{H^U_{33}}
\3tvec{u}{c}{t}_L \no \\
&+&\frac{1}{\sqrt{2}}(\bar{d},\bar{s},\bar{b})_R
\Mat3{Y^D_1(\phi_{d,3}+i\rho_{d,3})}{\eta Y^D_4s_{dL}(\phi'_{d,2}+i\rho'_{d,2})}{-\eta Y^D_4c_{dL}(\phi'_{d,2}+i\rho'_{d,2})}
{\eta Y^D_5s_{dR}(\phi'_{d,2}+i\rho'_{d,2})}{H^D_{22}}{H^D_{23}}
{-\eta Y^D_5c_{dR}(\phi'_{d,2}+i\rho'_{d,2})}{H^D_{32}}{H^D_{33}}
\3tvec{d}{s}{b}_L \no \\
&+&\frac{1}{\sqrt{2}}(\bar{e},\bar{\mu},\bar{\tau})_R
\Mat3{-Y^E_2(\phi_{d,3}+i\rho_{d,3})}{0}{0}
{0}{Y^E_3(\phi'_{d,1}+i\rho'_{d,1})}{-Y^E_3(\phi'_{d,2}+i\rho'_{d,2})}
{0}{Y^E_1(\phi'_{d,2}+i\rho'_{d,2})}{Y^E_1(\phi'_{d,1}+i\rho'_{d,1})}
\3tvec{e}{\mu}{\tau}_L +h.c. ,\\
H^U_{22}&=&[Y^U_1c_{uR}(\phi_{u,3}+i\rho_{u,3})-Y^U_5s_{uR}(\phi'_{u,1}+i\rho'_{u,1})]c_{uL} \no \\
&-&[Y^U_4c_{uR}(\phi'_{u,1}+i\rho'_{u,1})-|Y^U_3|s_{uR}(\phi_{u,3}+i\rho_{u,3})]s_{uL}, \\
H^U_{23}&=&[Y^U_1c_{uR}(\phi_{u,3}+i\rho_{u,3})-Y^U_5s_{uR}(\phi'_{u,1}+i\rho'_{u,1})]s_{uL} \no \\
&+&[Y^U_4c_{uR}(\phi'_{u,1}+i\rho'_{u,1})-|Y^U_3|s_{uR}(\phi_{u,3}+i\rho_{u,3})]c_{uL} , \\
H^U_{32}&=&[Y^U_1s_{uR}(\phi_{u,3}+i\rho_{u,3})+Y^U_5c_{uR}(\phi'_{u,1}+i\rho'_{u,1})]c_{uL} \no \\
&-&[Y^U_4s_{uR}(\phi'_{u,1}+i\rho'_{u,1})+|Y^U_3|c_{uR}(\phi_{u,3}+i\rho_{u,3})]s_{uL} , \\
H^U_{33}&=&[Y^U_1s_{uR}(\phi_{u,3}+i\rho_{u,3})+Y^U_5c_{uR}(\phi'_{u,1}+i\rho'_{u,1})]s_{uL} \no \\
&+&[Y^U_4s_{uR}(\phi'_{u,1}+i\rho'_{u,1})+|Y^U_3|c_{uR}(\phi_{u,3}+i\rho_{u,3})]c_{uL} ,\\
H^D_{22}&=&[Y^D_1c_{dR}(\phi_{d,3}+i\rho_{d,3})-\eta Y^D_5s_{dR}(\phi'_{d,1}+i\rho'_{d,1})]c_{dL} \no \\
&-&\eta [Y^D_4c_{dR}(\phi'_{d,1}+i\rho'_{d,1})-\eta |Y^D_3|s_{dR}(\phi_{d,3}+i\rho_{d,3})]s_{dL} ,\\
H^D_{23}&=&[Y^D_1c_{dR}(\phi_{d,3}+i\rho_{d,3})-\eta Y^D_5s_{dR}(\phi'_{d,1}+i\rho'_{d,1})]s_{dL} \no \\
&+&\eta [Y^D_4c_{dR}(\phi'_{d,1}+i\rho'_{d,1})-\eta |Y^D_3|s_{dR}(\phi_{d,3}+i\rho_{d,3})]c_{dL} , \\
H^D_{32}&=&[Y^D_1s_{dR}(\phi_{d,3}+i\rho_{d,3})+\eta Y^D_5c_{dR}(\phi'_{d,1}+i\rho'_{d,1})]c_{dL} \no \\
&-&\eta [Y^D_4s_{dR}(\phi'_{d,1}+i\rho'_{d,1})+\eta |Y^D_3|c_{dR}(\phi_{d,3}+i\rho_{d,3})]s_{dL} , \\
H^D_{33}&=&[Y^D_1s_{dR}(\phi_{d,3}+i\rho_{d,3})+\eta Y^D_5c_{dR}(\phi'_{d,1}+i\rho'_{d,1})]s_{dL} \no \\
&+&\eta [Y^D_4s_{dR}(\phi'_{d,1}+i\rho'_{d,1})+\eta |Y^D_3|c_{dR}(\phi_{d,3}+i\rho_{d,3})]c_{dL}, \\
\eta&=&e^{-i\gamma},
}
where
\eqn{
&&\phi'_{u,1}=\phi_{u,1}c_u+\phi_{u,2}s_u,\quad \rho'_{u,1}=\rho_{u,1}c_u+\rho_{u,2}s_u, \\
&&\phi'_{d,1}=\phi_{d,1}c_d+\phi_{d,2}s_d,\quad
\rho'_{d,1}=\rho_{d,1}c_d+\rho_{d,2}s_d. } From these
interactions, we can evaluate FCNC processes. For example, one can
see that $\phi'_{d,2}$ and $\rho'_{d,2}$ mediate flavor changing
operator such as $(\bar{d}_Rd_L)(\bar{d}_Ls_R)$ ,which contributes
$K^0-\bar{K}^0$ mass difference $\Delta m_K$. Note that the terms
$(\bar{d}_Rs_L)(\bar{d}_Rs_L)$ and $(\bar{d}_Ls_R)(\bar{d}_Ls_R)$
are not induced because contributions to them from $\phi'_{d,2}$
and $\rho'_{d,2}$ are cancelled due to degeneration of masses.
However, lepton flavor changing processes such as $\mu\to
e\gamma$, $\tau\to\mu\gamma$ and $\tau\to e\gamma$ are not
induced.

Flavor violating effective interactions are given by
\eqn{
{\cal L}_{Higgs-FCNC}&=&\frac{Y^U_4Y^U_5s_{uL}s_{uR}}{m^2_{\phi'_{u,2}}}
(\bar{u}_{R,\alpha}c_L^\alpha)(\bar{u}_{L,\beta}c_R^\beta)
+\frac{Y^D_4Y^D_5s_{dL}s_{dR}}{m^2_{\phi'_{d,2}}}
(\bar{d}_{R,\alpha}s_L^\alpha)(\bar{d}_{L,\beta}s_R^\beta) \no \\
&+&\frac{Y^D_4Y^D_5c_{dL}c_{dR}}{m^2_{\phi'_{d,2}}}
(\bar{d}_{R,\alpha}b_L^\alpha)(\bar{d}_{L,\beta}b_R^\beta) ,
}
where $\alpha$ and $\beta$ are color indices.
From this Lagrangian, we can evaluate the Higgs contributions to $\Delta m_K$,
$\Delta m_B$ and $\Delta m_D$
as follows:
\eqn{
(\Delta m_K)_{Higgs}&=&2Re\L<K^0|(-{\cal L}_{Higgs-FCNC})|\bar{K}^0\R> \no \\
&=&-2\frac{Y^D_4Y^D_5s_{dL}s_{dR}}{m^2_{\phi'_{d,2}}}
\L<K^0\L|(\bar{d}_{R,\alpha}s_L^\alpha)(\bar{d}_{L,\beta}s_R^\beta)\R|\bar{K}^0\R>, \\
(\Delta m_B)_{Higgs}&=&2Re\L<B^0|(-{\cal L}_{Higgs-FCNC})|\bar{B}^0\R> \no \\
&=&-2\frac{Y^D_4Y^D_5c_{dL}c_{dR}}{m^2_{\phi'_{d,2}}}
\L<B^0\L|(\bar{d}_{R,\alpha}b_L^\alpha)(\bar{d}_{L,\beta}b_R^\beta)\R|\bar{B}^0\R>, \\
(\Delta m_D)_{Higgs}&=&2Re\L<D^0|(-{\cal L}_{Higgs-FCNC})|\bar{D}^0\R> \no \\
&=&-2\frac{Y^U_4Y^U_5s_{uL}s_{uR}}{m^2_{\phi'_{u,2}}}
\L<D^0\L|(\bar{u}_{R,\alpha}c_L^\alpha)(\bar{u}_{L,\beta}c_R^\beta)\R|\bar{D}^0\R>,
}
where
\eqn{
\L<K^0\L|(\bar{d}_{R,\alpha}s_L^\alpha)(\bar{d}_{L,\beta}s_R^\beta)\R|\bar{K}^0\R>
&=&\L[\frac{1}{24}+\frac14\L(\frac{m_{K^0}}{m_s(2GeV)+m_d(2GeV)}\R)^2\R]m_{K^0}f^2_K \no \\
&=&6.56\times 10^7 MeV^3, \\
\L<B^0\L|(\bar{d}_{R,\alpha}b_L^\alpha)(\bar{d}_{L,\beta}b_R^\beta)\R|\bar{B}^0\R>
&=&\L[\frac{1}{24}+\frac14\L(\frac{m_{B^0}}{m_b(m_b)+m_d(m_b)}\R)^2\R]m_{B^0}f^2_B
=9.21\times 10^7 MeV^3, \\
\L<D^0\L|(\bar{u}_{R,\alpha}c_L^\alpha)(\bar{u}_{L,\beta}c_R^\beta)\R|\bar{D}^0\R>
&=&\L[\frac{1}{24}+\frac14\L(\frac{m_{D^0}}{m_c(m_c)+m_u(m_c)}\R)^2\R]m_{D^0}f^2_D
=4.99\times 10^7 MeV^3,
}
which are evaluated by using the parameters given in appendix.
Requiring $|(\Delta m_M)_{Higgs}|< \Delta m_M (M=K,B,D)$, we get
\eqn{
&&m_{\phi'_{d,2}}>4.6TeV\quad (\Delta m_K), \\
&&m_{\phi'_{d,2}}>4.0TeV\quad (\Delta m_B), \\
&&m_{\phi'_{u,2}}>37.6TeV\quad (\Delta m_D). } 
These constraints
are too strong. In our model, SUSY contributions to FCNC may be used to
cancel these Higgs contributions. However, in order to suppress
$\Delta m_{K,B,D}$, we must give three cancellation conditions:
\eqn{
|(\Delta m_M)_{Higgs}+(\Delta m_M)_{SUSY}| \ll |(\Delta m_M)_{Higgs}| \quad (M=K,B,D),
}
which is unnatural.
In the next subsection, we show the number of  cancellation conditions are reduced to two from three.

\subsection{Squark and gluino contributions}

Now we evaluate SUSY-FCNC contributions. 
As we assume $\Delta m_D$  is suppressed by cancellation:
\eqn{
|(\Delta m_D)_{Higgs}+(\Delta m_D)_{SUSY}| \ll |(\Delta m_D)_{Higgs}|,
}
we consider only $K^0-\bar{K}^0$ and $B^0-\bar{B}^0$
mass differences induced by squark and gluino box diagrams. These
contributions depend only on down type squark mass matrices.
Considering the following squark Lagrangian \eqn{
-{\cal L}_{squark}&=&m^2_Q(|Q_1|^2+|Q_2|^2)+m^2_{Q_3}|Q_3|^2
+m^2_D(|D^c_1|^2+|D^c_2|^2)+m^2_{D_3}|D^c_3|^2 \no \\
&+&\L\{e^{-i\phi_Q}|m^2_{BQ}|Q^\dagger_3(c_QQ_1+s_QQ_2)
+e^{i\phi_D}|m^2_{BD}|(D^c_3)^\dagger(c_DD^c_1+s_DD^c_2)+h.c. \R\} \no \\
&+&(D-terms) ,
}
one can see that down type squark mass matrix is given by
\eqn{
-{\cal L}_{down-squark}&=&(D^\dagger, D^c)
\mat2{M^2_{LL}}{0}{0}{M^2_{RR}}
\2tvec{D}{(D^c)^\dagger}, \\
M^2_{LL}&=&\Mat3{m^2_Q}{0}{e^{i\phi_Q}|m^2_{BQ}|c_Q}
{0}{m^2_Q}{e^{i\phi_Q}|m^2_{BQ}|s_Q}
{e^{-i\phi_Q}|m^2_{BQ}|c_Q}{e^{-i\phi_Q}|m^2_{BQ}|s_Q}{m^2_{Q_3}}, \\
M^2_{RR}&=&\Mat3{m^2_D}{0}{e^{i\phi_D}|m^2_{BD}|c_D}
{0}{m^2_D}{e^{i\phi_D}|m^2_{BD}|s_D}
{e^{-i\phi_D}|m^2_{BD}|c_D}{e^{-i\phi_D}|m^2_{BD}|s_D}{m^2_{D_3}}
. } Where D-term contributions are absorbed into
$m^2_{Q,Q_3,D,D_3}$. In super-CKM basis, squark mass matrices are
given by \eqn{
(M^2_{LL})'&=&V^\dagger_{dL} M^2_{LL}V_{dL} \no \\
&=&\Mat3{m^2_Q}{\eta_Q |m^2_{BQ}|s_{Qd}s_{dL}}{-\eta_Q |m^2_{BQ}|s_{Qd}c_{dL}}
{\eta^*_Q |m^2_{BQ}|s_{Qd}s_{dL}}{(M^2_{Ld})_{22}}{(M^2_{Ld})_{23}}
{-\eta^*_Q |m^2_{BQ}|s_{Qd}c_{dL}}{(M^2_{Ld})^*_{23}}{(M^2_{Ld})_{33}}, \\
(M^2_{Ld})_{22}&=&m^2_Q(c_{dL})^2+m^2_{Q_3}(s_{dL})^2-|m^2_{BQ}|c_{Qd}s_{dL}c_{dL}(\eta_Q+\eta^*_Q) , \\
(M^2_{Ld})_{23}&=&(m^2_Q-m^2_{Q_3})c_{dL}s_{dL}+|m^2_{BQ}|c_{Qd}(\eta_Q(c_{dL})^2-\eta^*_Q (s_{dL})^2) , \\
(M^2_{Ld})_{33}&=&m^2_Q(s_{dL})^2+m^2_{Q_3}(c_{dL})^2+|m^2_{BQ}|c_{Qd}s_{dL}c_{dL}(\eta_Q + \eta^*_Q) , \\
s_{Qd}&=&\sin(\theta_Q-\theta_d) , \\
\eta_Q&=&\eta e^{i\phi_Q}, \\
(M^2_{RR})'&=&V^\dagger_{dR} M^2_{RR}V_{dR} \no \\
&=&\Mat3{m^2_D}{\eta_D |m^2_{BD}|s_{Dd}s_{dR}}{-\eta_D |m^2_{BD}|s_{Dd}c_{dR}}
{\eta^*_D |m^2_{BD}|s_{Dd}s_{dR}}{(M^2_{Rd})_{22}}{(M^2_{Rd})_{23}}
{-\eta^*_D |m^2_{BD}|s_{Dd}c_{dR}}{(M^2_{Rd})^*_{23}}{(M^2_{Rd})_{33}} , \\
(M^2_{Rd})_{22}&=&m^2_D(c_{dR})^2+m^2_{D_3}(s_{dR})^2-|m^2_{BD}|c_{Dd}s_{dR}c_{dR}(\eta_D+\eta^*_D),  \\
(M^2_{Rd})_{23}&=&(m^2_D-m^2_{D_3})c_{dR}s_{dR}+|m^2_{BD}|c_{Dd}(\eta_D(c_{dR})^2-\eta^*_D(s_{dR})^2),  \\
(M^2_{Rd})_{33}&=&m^2_D(s_{dR})^2+m^2_{D_3}(c_{dR})^2+|m^2_{BD}|c_{Dd}s_{dR}c_{dR}(\eta_D+\eta^*_D),  \\
s_{Dd}&=&\sin(\theta_D-\theta_d),\\
\eta_D&=&\eta^* e^{i\phi_D}. } Here we assume degenerated mass
squared parameters as \eqn{ m^2_Q=m^2_{Q_3},\quad m^2_D=m^2_{D_3},
} which are essential assumptions to realize cancellation between
Higgs and SUSY-FCNC contributions. These relations are realized if
gaugino mass contributions dominate in RGEs. With this assumption,
diagonal elements of mass squared matrix are also degenerated
approximately  as follows: \eqn{
&&(M^2_{Ld})_{22}\simeq (M^2_{Ld})_{33}\simeq m^2_Q, \\
&&(M^2_{Rd})_{22}\simeq (M^2_{Rd})_{33}\simeq m^2_D. } Where we
assume that the contributions from $m^2_{BD,BQ}$ are negligible.
Furthermore, we assume $\eta_Q=\eta_D=1$ to suppress \eqn{
Im\L<K^0|{\cal L}_{SUSY-FCNC}|\bar{K}^0\R>. } Here flavor changing
effective interactions induced by squark and gluino box diagrams
are calculated in mass insertion approximation \cite{box} as
follows: \eqn{ {\cal
L}_{SUSY-FCNC}&=&\frac{\alpha^2_3}{216M^2_{Q,K}}
\L\{(\delta_{12})^2_{LL}\L[24xf_1(x)+66f_2(x)\R]O_1
+(\delta_{12})^2_{RR}\L[24xf_1(x)+66f_2(x)\R]O_2 \R. \no \\
&+&\L.  (\delta_{12})_{LL}(\delta_{12})_{RR}
\L[(504xf_1(x)-72f_2(x))O_3+(24xf_1(x)+120f_2(x))O_4\R] \R\} \no \\
&+&\frac{\alpha^2_3}{216M^2_{Q,B}}
\L\{(\delta_{13})^2_{LL}\L[24yf_1(y)+66f_2(y)\R]P_1
+(\delta_{13})^2_{RR}\L[24yf_1(y)+66f_2(y)\R]P_2 \R. \no \\
&+&\L.  (\delta_{13})_{LL}(\delta_{13})_{RR}
\L[(504yf_1(y)-72f_2(y))P_3+(24yf_1(y)+120f_2(y))P_4\R] \R\}, }
where $\alpha_3$ is $SU(3)_c$ gauge coupling, $M_{Q,K}$ and
$M_{Q,B}$ are averaged squark mass, and the
other parameters are defined as \eqn{
&&f_1(x)=\frac{6(1+3x)\ln x+x^3-9x^2-9x+17}{6(x-1)^5}, \\
&&f_2(x)=\frac{6x(1+x)\ln x-x^3-9x^2+9x+1}{3(x-1)^5}, \\
&&O_1=(\bar{d}_{L,\alpha}\gamma_\mu s_L^\alpha)(\bar{d}_{L,\beta}\gamma^\mu s_L^\beta), \quad
P_1=(\bar{d}_{L,\alpha}\gamma_\mu b_L^\alpha)(\bar{d}_{L,\beta}\gamma^\mu b_L^\beta), \\
&&O_2=(\bar{d}_{R,\alpha}\gamma_\mu s_R^\alpha)(\bar{d}_{R,\beta}\gamma^\mu s_R^\beta), \quad
P_2=(\bar{d}_{R,\alpha}\gamma_\mu b_R^\alpha)(\bar{d}_{R,\beta}\gamma^\mu b_R^\beta), \\
&&O_3=(\bar{d}_{R,\alpha}s_L^\alpha)(\bar{d}_{L,\beta}s_R^\beta), \quad
P_3=(\bar{d}_{R,\alpha}b_L^\alpha)(\bar{d}_{L,\beta}b_R^\beta), \\
&&O_4=(\bar{d}_{R,\alpha}s_L^\beta)(\bar{d}_{L,\beta}s_R^\alpha), \quad
P_4=(\bar{d}_{R,\alpha}b_L^\beta)(\bar{d}_{L,\beta}b_R^\alpha), \\
&&(\delta_{12})_{LL}=\frac{|m^2_{BQ}|s_{Qd}s_{dL}}{M^2_{Q,K}} , \quad
(\delta_{13})_{LL}=\frac{-|m^2_{BQ}|s_{Qd}c_{dL}}{M^2_{Q,B}}, \\
&&(\delta_{12})_{RR}=\frac{|m^2_{BD}|s_{Dd}s_{dR}}{M^2_{Q,K}}, \quad
(\delta_{13})_{RR}=\frac{-|m^2_{BD}|s_{Dd}c_{dR}}{M^2_{Q,B}}, \\
&&x=\frac{M^2_3}{M^2_{Q,K}},\quad y=\frac{M^2_3}{M^2_{Q,B}},}where
$M_3$ is gluino mass.
With the assumptions of Eqs.(150) and (151), one can assume
$M^2_{Q,K}=M^2_{Q,B}$. Note that the dominant contributions to
$\Delta m_K$ and $\Delta m_B$ come from $O_3$ and $P_3$ in
Eq.(153) due to the large coefficients. Total contributions to
$O_3$ and $P_3$ from Higgs (Eq.(121)) and SUSY (Eq.(153)) are
written as follows: \eqn{ {\cal
L}_{O_3}&=&\L[\frac{Y^D_4Y^D_5}{m^2_{\phi'_{d,2}}}
+\frac{\alpha^2_3
|m^2_{BQ}m^2_{BD}|s_{Qd}s_{Dd}}{216M^6_{Q,K}}(504xf_1(x)-72f_2(x))\R]s_{dL}s_{dR}
(\bar{d}_{R,\alpha}s_L^\alpha)(\bar{d}_{L,\beta}s_R^\beta),  \\
{\cal L}_{P_3}&=&\L[\frac{Y^D_4Y^D_5}{m^2_{\phi'_{d,2}}}
+\frac{\alpha^2_3
|m^2_{BQ}m^2_{BD}|s_{Qd}s_{Dd}}{216M^6_{Q,K}}(504xf_1(x)-72f_2(x))\R]c_{dL}c_{dR}
(\bar{d}_{R,\alpha}b_L^\alpha)(\bar{d}_{L,\beta}b_R^\beta). } If
accidental cancellation occurs between the terms in bracket
$[\quad]$, new physics contributions to $\Delta m_K$ and $\Delta
m_B$ are well-suppressed at same time. Assuming $x=1$ and
$|m_{BQ}|s_{Qd}=-2|m_{BD}|s_{Dd}$, we get \eqn{
f_1(1)=\frac{1}{20},\quad f_2(1)=-\frac{1}{30}, } and cancellation
condition: \eqn{ \frac{Y^D_4Y^D_5}{m^2_{\phi'_{d,2}}}
=13.8\frac{\alpha^2_3(|m^2_{BQ}|s_{Qd})^2}{216M^6_{Q,K}}. }

One finds that Eq.(166) is satisfied, for example, if we put \eqn{
\alpha_3=0.12,\quad m^2_{\phi'_{d,2}}=M^2_{Q,K},\quad
\frac{|m^2_{BQ}s_{Qd}|}{M^2_{Q,K}}=0.218. } Then the sub-dominant
contributions from Eq.(153) are evaluated as follows: \eqn{
(\Delta m_K)_{SUSY}&=&-\frac{Y^D_4Y^D_5}{13.8M^2_{Q,K}} \times
2Re\L<K^0\L|\L\{-\L[s^2_{dL}O_1+\frac14s^2_{dR}O_2\R]
+\frac{2.8}{2}s_{dL}s_{dR}O_4\R\}\R|\bar{K}^0\R> \no \\
&=&1.06\times 10^{-12}\L(\frac{TeV}{M_{Q,K}}\R)^2 MeV , \\
(\Delta m_B)_{SUSY}&=&-\frac{Y^D_4Y^D_5}{13.8M^2_{Q,K}}
\times 2Re\L<B^0\L|\L\{-\L[c^2_{dL}P_1+\frac14 c^2_{dR}P_2\R]+\frac{2.8}{2}c_{dR}c_{dL}P_4\R\}\R|\bar{B}^0\R> \no \\
&=&1.74\times 10^{-10}\L(\frac{TeV}{M_{Q,K}}\R)^2 MeV  ,
}
where
\eqn{
&&\L<K^0|O_1|\bar{K}^0\R>=\L<K^0|O_2|\bar{K}^0\R>=\frac13 m_Kf^2_K=4.25\times 10^6MeV^3, \\
&&\L<K^0|O_4|\bar{K}^0\R>=\L[\frac18+\frac{1}{12}\L(\frac{m_{K^0}}{m_s(2GeV)+m_d(2GeV)}\R)^2\R]m_Kf^2_K
=2.33\times 10^7MeV^3,\\
&&\L<B^0|P_1|\bar{B}^0\R>=\L<B^0|P_2|\bar{B}^0\R>=\frac13 m_Bf^2_B=7.04\times 10^7MeV^3, \\
&&\L<B^0|P_4|\bar{B}^0\R>=\L[\frac18+\frac{1}{12}\L(\frac{m_{B^0}}{m_b(m_b)+m_d(m_b)}\R)^2\R]m_Bf^2_B
=5.41\times 10^7MeV^3,
}
are used. Requiring $|(\Delta m_M)_{SUSY}|< \Delta m_M (M=K,B)$, we get
\eqn{
&&M_{Q,K}=m_{\phi'_{d,2}}>0.6TeV\quad (\Delta m_K), \\
&&M_{Q,K}=m_{\phi'_{d,2}}>0.7TeV\quad (\Delta m_B). } One finds
that these constraints are weaker than Eqs. (128) and (129).
Therefore three cancellation conditions Eq.(131) are reduced to two conditions
Eq.(132) and Eq.(166).


\section{Summary}

In this paper, we have considered the Higgs-FCNC problem in
$S_4\times Z_2$ flavor symmetric
extra U(1) model, and have shown that
the Higgs mass bounds from FCNCs are weaken by cancellation between Higgs and SUSY contributions.
As the result, SUSY breaking scale may be around $O(TeV)$ region.
It might be expected that the new gauge symmetry
 and the flavor symmetry are tested in LHC or future colliders.

\section*{Acknowledgments}
H.O. thanks great hospitality of the Kanazawa Institute for
Theoretical Physics at Kanazawa University. Discussions during my
visit were fruitful to finalize this project.
H. O. acknowledges partial support from the Science and Technology Development Fund (STDF)
project ID 437 and the ICTP project ID 30.

\appendix

\section*{Appendix}

\section{Experimental Values}

Running masses of quarks and charged leptons \cite{mass}:
\eqn{
\begin{tabular}{lll}
$m_u(m_Z)=1.28^{+0.50}_{-0.39} ({\rm MeV})$, &
$m_c(m_Z)=624\pm 83\ ({\rm MeV})$, &
$m_t(m_Z)=172.5\pm 3.0 ({\rm GeV})$, \\
$m_d(m_Z)=2.91^{+1.24}_{-1.20}\ ({\rm MeV})$, &
$m_s(m_Z)=55^{+16}_{-15}\ ({\rm MeV})$, &
$m_b(m_Z)=2.89\pm 0.09\ ({\rm GeV})$, \\
$m_e(m_Z)=0.48657\ ({\rm MeV})$, &
$m_\mu(m_Z)=102.72\ ({\rm MeV})$, &
$m_\tau(m_Z)=1746\ ({\rm MeV})$.
\end{tabular}
}
CKM matrix elements and Jarlskog invariant \cite{PDG2008}:
\eqn{
&&
\begin{tabular}{lll}
$\L| V_{ud}\R|=0.97418\pm0.00027$, &
$\L| V_{us}\R|=0.2255\pm0.0019$,   &
$\L| V_{ub}\R|=(3.93\pm0.36)\times 10^{-3}$, \\
$\L| V_{cd}\R|=0.230\pm0.011$, &
$\L| V_{cs}\R|=1.04\pm0.06$,   &
$\L| V_{cb}\R|=(41.2\pm1.1)\times 10^{-3}$, \\
$\L| V_{td}\R|=(8.1\pm0.6)\times10^{-3}$, &
$\L| V_{ts}\R|=(38.7\pm2.3)\times10^{-3}$,&
$\L| V_{tb}\R|>0.74$,
\end{tabular} \no \\
&&J_{CP}=Im(V_{ud}V^*_{ub}V_{cd}V^*_{cb})=(3.05^{+0.19}_{-0.20})\times10^{-5}.
}
Neutrino mass-squared differences and the parameters of MNS matrix \cite{PDG2008}:
\begin{eqnarray}
&&\Delta m^2_{21}= m^2_{\nu_2}-m^2_{\nu_1}=(8.0\pm 0.3)\times 10^{-5} \quad (eV^2), \quad
\Delta m^2_{32} = \L|m^2_{\nu_3}-m^2_{\nu_2}\R|=(1.9-3.0)\times 10^{-3}\quad (eV^2), \no \\
&&V_{MNS}=\Mat3{c_{12}c_{13}}{s_{12}c_{13}}{s_{13}e^{-i\delta'}}
{-s_{12}c_{23}-c_{12}s_{23}s_{13}e^{i\delta'}}{c_{12}c_{23}-s_{12}s_{23}s_{13}e^{i\delta'}}{s_{23}c_{13}}
{s_{12}s_{23}-c_{12}c_{23}s_{13}e^{i\delta'}}{-c_{12}s_{23}-s_{12}c_{23}s_{13}e^{i\delta'}}{c_{23}c_{13}}
\Mat3{1}{0}{0}{0}{e^{i\alpha'}}{0}{0}{0}{e^{i\beta'}}, \no \\
&&\theta_{12}=34.0^\circ {}^{+1.3^\circ}_{-1.5^\circ}, \quad
45.0^\circ > \theta_{23}>36.8^\circ, \quad
12.9^\circ >\theta_{13}>0.0^\circ.
\end{eqnarray}
Meson masses \cite{PDG2008}: \eqn{ m_{K^0}=497.614\pm
0.024MeV,\quad m_{B^0}=5279.50\pm 0.30MeV ,\quad
m_{D^0}=1864.84\pm 0.17MeV. } Meson mass differences
\cite{PDG2008}: \eqn{
&&\Delta m_K=(3.483\pm 0.006)\times 10^{-12}MeV, \no \\
&&\Delta m_B=(3.337\pm 0.033)\times 10^{-10}MeV, \no \\
&&\Delta m_D=(1.56\pm 0.43)\times 10^{-11}MeV. } Meson decay
constants \cite{decay-const}: \eqn{ f_K=159.8\pm 1.4 \pm 0.44
MeV,\quad f_B=200\pm 20 MeV,\quad f_D=212 \pm 14 MeV. } Running
quark mass \cite{mass}: \eqn{
&&m_d(2GeV)=5.04^{+0.96}_{-1.54}MeV,\quad m_s(2GeV)=105^{+25}_{-35}MeV, \no \\
&&m_d(m_b)=4.23^{+1.74}_{-1.71} MeV ,\quad m_b(m_b)=4.20\pm 0.07 GeV, \no \\
&&m_u(m_c)=2.57^{+0.99}_{-0.84}MeV, \quad m_c(m_c)=1.25\pm 0.09GeV.
}




\begin{thebibliography}{99}
\bibitem{SUSY}H.P. Nilles,  \PRP\vol{110}{1984}{1}.
\bibitem{extra-u1}F. Zwirner,  \IJMP\vol{A3}{1988}{49},
J.L. Hewett and T.G. Rizzo,  \PRP\vol{183}{1989}{193}.
\bibitem{mu-problem}D. Suematsu and Y. Yamagishi,  \IJMP\vol{A10}{1995}{4521}.
\bibitem{f-extra-u1}R. Howl and S.F. King, JHEP\vol{0805}{2008}{008} [arXiv:0802.1909[hep-ph]].
\bibitem{review}
For reviews and recent advanced works, see the followings:
  P.~O.~Ludl,
  arXiv:0907.5587 [hep-ph].
  H.~Ishimori, T.~Kobayashi, H.~Ohki, H.~Okada, Y.~Shimizu and M.~Tanimoto,
  arXiv:1003.3552 [hep-th],
  L.~Merlo,
  arXiv:1004.2211 [hep-ph],
  W.~Grimus and P.~O.~Ludl,
  arXiv:1006.0098 [hep-ph],
  K.~S.~Babu and S.~Gabriel,
  arXiv:1006.0203 [hep-ph],
  P.~O.~Ludl,
  arXiv:1006.1479 [math-ph],
 T. J. Burrows and S. F. King,  arXiv:1007.2310 [hep-ph]
S. Forste, H. P. Nilles, S. Ramos-Sanchez and P. K.S. Vaudrevange,  arXiv:1007.3915 [hep-ph],
J. Barry and W. Rodejohann, arXiv:1007.5217 [hep-ph].

\bibitem{s4}
S. Pakvasa and H. Sugawara,  \PL\vol{B73}{1978}{61},
E. Ma, \PL\vol{B632}{2006}{352} [arXiv:hep-ph/0508231],
C. Hagedorn, M. Lindner and R.N. Mohapatra, JHEP\vol{0606}{2006}{042}
[arXiv:hep-ph/0602244],
 H.~Zhang,
  Phys.\ Lett.\  B {\bf 655}, 132 (2007)
  [arXiv:hep-ph/0612214],
Y. Koide, JHEP\vol{0708}{2007}{086}
[arXiv:0705.2275 [hep-ph]],
F. Bazzocchi and S. Morisi, [arXiv:0811.0345 [hep-ph]],
H.~Ishimori, Y.~Shimizu and M.~Tanimoto,
  Prog.\ Theor.\ Phys.\  {\bf 121}, 769 (2009)
  [arXiv:0812.5031 [hep-ph]],
F.~Bazzocchi, L.~Merlo and S.~Morisi,
  Nucl.\ Phys.\  B {\bf 816} (2009) 204
  [arXiv:0901.2086 [hep-ph]],
F.~Bazzocchi, L.~Merlo and S.~Morisi,
  Phys.\ Rev.\  D {\bf 80} (2009) 053003
  [arXiv:0902.2849 [hep-ph]],
   G.~Altarelli, F.~Feruglio and L.~Merlo,
  JHEP {\bf 0905} (2009) 020
  [arXiv:0903.1940 [hep-ph]],
  W.~Grimus, L.~Lavoura and P.~O.~Ludl,
  J.\ Phys.\ G {\bf 36}, 115007 (2009)
  [arXiv:0906.2689 [hep-ph]],
  G.~J.~Ding,
  Nucl.\ Phys.\  B {\bf 827}, 82 (2010)
  [arXiv:0909.2210 [hep-ph]],
   B.~Dutta, Y.~Mimura and R.~N.~Mohapatra,
  arXiv:0911.2242 [hep-ph],
  D.~Meloni,
  J.\ Phys.\ G {\bf 37}, 055201 (2010)
  [arXiv:0911.3591 [hep-ph]],
  G.~J.~Ding and J.~F.~Liu,
  JHEP {\bf 1005}, 029 (2010)
  [arXiv:0911.4799 [hep-ph]],
  S.~Morisi and E.~Peinado,
  Phys.\ Rev.\  D {\bf 81}, 085015 (2010)
  [arXiv:1001.2265 [hep-ph]],
 C.~Hagedorn, S.~F.~King and C.~Luhn,
  JHEP {\bf 1006}, 048 (2010)
  [arXiv:1003.4249 [hep-ph]],
  Y.~H.~Ahn, S.~K.~Kang, C.~S.~Kim and T.~P.~Nguyen,
  arXiv:1004.3469 [hep-ph],
  H.~Ishimori, K.~Saga, Y.~Shimizu and M.~Tanimoto,
  Phys.\ Rev.\  D {\bf 81}, 115009 (2010)
  [arXiv:1004.5004 [hep-ph]],
R. N. Mohapatra,  arXiv:1007.1633 [hep-ph],
  P.~V.~Dong, H.~N.~Long, D.~V.~Soa and V.~V.~Vien,
  arXiv:1009.2328 [hep-ph].


\bibitem{a4}E.Ma and G.Rajasekaran, \PR\vol{D64}{2001}{113012} [arXiv:hep-ph/0106291],
K.S. Babu, E. Ma and J.W.F. Valle, \PL\vol{B552}{2003}{207} [arXiv:hep-ph/0206292],
G. Altarelli and F. Feruglio, \NP\vol{B741}{2006}{215} [arXiv:hep-ph/0512103],
X.G. He, Y.Y. Keum and R.R. Volkas, JHEP\vol{0604}{2006}{039} [arXiv:hep-ph/0601001],
M. Honda and M. Tanimoto, \PTP\vol{119}{2008}{583} [arXiv:0801.0181 [hep-ph]].
\bibitem{t-prime}A. Aranda, C.D. Carone and R.F. Lebed, \PR\vol{D62}{2000}{016009} [arXiv:hep-ph/0002044],
P.H. Frampton and T.W. Kephart, JHEP\vol{0709}{2007}{110}
[arXiv:0706.1186 [hep-ph]],
S. Sen, \PR\vol{D76}{2007}{115020} [arXiv:0710.2734 [hep-ph]],
G.J. Ding, \PR\vol{D78}{2008}{036011}
[arXiv:0803.2278 [hep-ph]],
M.C. Chen, K.T. Mahanthappa and F. Yu, arXiv:0907.3963 [hep-ph],
 F.~Feruglio, C.~Hagedorn, Y.~Lin and L.~Merlo,
  Nucl.\ Phys.\  B {\bf 775} (2007) 120
  [arXiv:hep-ph/0702194].
\bibitem{d27}I.de Medeiros Varzielas, S.F. King and G.G. Ross,
\PL\vol{B648}{2007}{201} [arXiv:hep-ph/0607045], E. Ma,
\MPL\vol{A21}{2006}{1917} [arXiv:hep-ph/0607056],
E. Ma,
  Phys.\ Lett.\  B {\bf 660}, 505 (2008)
  [arXiv:0709.0507 [hep-ph]].


\bibitem{d54}H. Ishimori, T. Kobayashi, H. Okada, Y. Shimizu and M. Tanimoto, JHEP \vol{0904}{2009}{011} [arXiv:0811.4683 [hep-ph]],
 H.~Ishimori, T.~Kobayashi, H.~Okada, Y.~Shimizu and M.~Tanimoto,
  JHEP {\bf 0912}, 054 (2009)
  [arXiv:0907.2006 [hep-ph]].
\bibitem{s4e6}Y.~Daikoku and H.~Okada, [arXiv:0910.3370](accepted for publication in PRD).

\bibitem{e6fcnc}B. A. Campbell, J. Ellis, K. Enqvist, M. K. Gaillard and D. V. Nanopoulos,
\IJMP\vol{A2}{1987}{831}

\bibitem{lepton}J. Kubo, \PL\vol{B578}{2004}{156}.

\bibitem{box}F. Gabbiani, E. Gabrielli, A. Masiero and L. Silvestrini,
\NP\vol{B477}{1996}{321}.
\bibitem{mass}Z. z. Xing, H. Zhang and S. Zhou, \PR\vol{D77}{2008}{113016} [arXiv:0712.1419[hep-ph]].
\bibitem{PDG2008}C. Amsler et al. (Particle Data Group), Phys. Lett. B667, 1 (2008) and
2009 partial update for the 2010 edition.
Cut-off date for this update was January 15, 2009.
\bibitem{decay-const}V. Lubicz and C. Tarantino, Nouvo Cim. {\bf 123B}(2008)674 [arXiv: 0807.4605[hep-lat]].
\end{thebibliography}
\end{document}